\documentclass[11pt]{article}
\usepackage{graphicx}
\usepackage{amssymb}
\title{Some First Steps Towards a Radiation GRMHD Code: \\
Radiative Effects on Accretion Rate onto a Kerr Black Hole}
\author{Jean-Pierre De Villiers\\
Calgary, Alberta\\
email: jpd\_asph@mac.com}
\begin{document}
\maketitle

\begin{center}
{\large {\bf Abstract}}
\end{center}

The role of radiation in general relativistic magnetohydrodynamic (GRMHD) accretion simulations
is discussed through axisymmetric simulations of the evolution of
an initial torus seeded with a weak magnetic field. The paper compares
and contrasts the rate of accretion onto a Kerr black hole and mass
flux out out of the initial torus at large radii 
 in the GRMHD code of De Villiers \& Hawley and a
newly developed radiative GRMHD code. This rGRMHD code currently
operates in the diffusion approximation, restricting the study of
radiative effects to the bound portion of the accretion disk/jet system.
However, these preliminary findings suggest that radiative effects do play
a potentially significant role in regulating the accretion flow. 

\section{Overview}

The task of simulating accretion disks in black hole systems has seen
much progress in recent years with the advent of general relativistic
MHD (GRMHD) codes capable of simulating disks under not only under axisymmetry (Gammie, McKinney, \& Toth, 2003) but also in three
spatial dimensions (De Villiers \& Hawley, 2003, hereafter DH03; and Mizuno {\it et al.}, 2006). In addition, the extraction 
of observables from
such simulations has been described in the literature, with 
ray tracing techniques applied in a post-processing phase to numerical
data generated by the GRMHD solver by Schnittman, Krolik, \& Hawley (2006). However, it is understood that radiative
processes can play a potentially significant role in the dynamics of 
black hole/disk/jet systems (Frank, King, \& Raine, 2003). So, ideally, the task of simulating such systems should be undertaken with codes that capture not only the relativistic dynamics of the accreting gas and magnetic fields, but also treat the radiation field in 
a self-consistent manner. Results of non-relativistic radiative shearing box simulations have been
reported by Hirose, Krolik, \& Stone (2006), showing that radiative
effects are important contributors to the dynamics of accretion flows. However, fully relativistic treatments of the radiation problem remain a rarity, even though discussion of the
formalism can be found in literature spanning
several decades, from Lindquist (1966), through the standard reference by Mihalas \& Weibel-Mihalas (1984; hereafter MM84), and recently in Takahashi (2007; hereafter T07), where a derivation of the equations of radiative hydrodynamics for the Kerr metric
are given.

The objective of the project introduced here is to create a radiative GRMHD code to explore the effects of radiation on accretion disk simulations.
The starting point is the GRMHD code of DH03, an explicit, finite difference solver that evolves the equations of general relativistic magnetohydrodynamics in the Kerr metric. Over a period of a few years the GRMHD code has been augmented by the
addition of a ray tracing module which forms part of the code's history
mechanism. This module can produce fully ray-traced, time-ordered 2D pixel maps
of emission processes captured as the simulation evolves; since the GRMHD code is a parallel MPI code, this extra component has been delegated to a slave processor so that the overall parallel efficiency of the original solver has
not significantly degraded (though execution times are slower). 
More recently, a radiation component has been added to the code's main physics loop, introducing the calculation of radiative effects to the source and transport steps that evolve the equations of
energy and momentum conservation. Although these additions have greatly increased the complexity
of the code, the ever-increasing power of desktop computing
systems allows axisymmetric simulations to be readily carried out, and even modest 3D simulations are within the reach of such systems (though 
with turnaround times measured in weeks, if not months). Large-scale simulations 
with this new rGRMHD code remain in the realm of massively parallel 
high-performance computing systems, and lie beyond the scope of the
project described here.

The addition of radiation to the energy/momentum equations represents
only one of two major modifications to the rGRMHD code. The second is the introduction
of the radiative transfer equation (RTE) to correctly capture the flow of radiation through the
simulation volume; the addition of the RTE is an ongoing project. Since the modification
of the dynamical equations is itself a significant change, it is desirable to test the code as it stands prior to investing further development time on the RTE. As it turns out, it is possible to decouple
the dynamical equations from the RTE by working in the diffusion approximation, where
the radiating fluid is assumed to be sufficiently opaque that the radiation field is thermal (MM84).
The main computational advantage of the diffusion approximation is that the radiative effects can be obtained
from fluid variables given the availability of a suitable opacity model.
This approximation
does, however, impose a set of restrictions since only certain portions of the
simulation volume will satisfy its underlying conditions. The simulations
described here study radiative effects in the bound portion of the accretion
flow only, where the fluid is sufficiently dense. Radiative effects are switched off in the
unbound regions, namely in the evacuated funnel where fast outflows
and jets have been reported by De Villiers, Hawley, \& Krolik (2003; hereafter DHK03). Since radiative effects in the dynamics of the outflows and
jets are thought to be significant, the results of the simulations described here should be
treated as informative, but certainly not definitive or comprehensive.

To perform this preliminary assessment of the
importance of radiative effects on accretion, two families of axisymmetric simulations
loosely based on model KDP of DHK03 are used to 
compare mass fluxes for different temperatures in the initial tori
and two equations of state.
In an effort to prevent the technical details from distracting from the
results, a description of the present state of the rGRMHD code has been relegated to the 
appendices, and the emphasis of the body of the paper is on the
simulations.

\section{Initial State and Boundary Conditions}

The earlier accretion studies of DHK03 and follow-on papers (Hirose {\it et al}, 2004;
De Villiers {\it et al.}, 2005; Krolik, Hawley, \& Hirose, 2005)
operated under the test fluid approximation, meaning that the mass-energy 
of the initial torus was effectively decoupled from the mass of
the central black hole, so that simulations could equally be interpreted for
accretion flows in systems of stellar or galactic-core scale. In incorporating 
radiative effects, this arbitrary scaling is no longer possible
and the density and temperature of
the initial torus must be calibrated to `realistic' astrophysical
conditions.

In constructing an initial state for the rGRMHD code, it is
necessary to discretize an initial torus on a grid that provides both good spatial resolution and rapid numerical execution. These two constraints lead to the choice of a compact initial torus that has its pressure maximum very
close to the black hole (hence the use of quotation marks in the preceding
paragraph). Since the rGRMHD code uses extremal light-crossing time to set the time step size (DH03), no additional constraints are
imposed by the addition of radiation. Radiative simulations are slower
than their non-radiative counterparts, but this is only due to added computational complexity and not to overly restrictive time step sizes. The equations of
the initial state are derived in DHK03 and summarized in Appendix \ref{InitState}. 

The general simulation parameters are listed in Table \ref{simparams}.
As in earlier work, the grid is chosen to have 192 radial and polar zones, and one azimuthal zone (i.e. a $2.5\,D$ simulation).  To make 
contact with earlier KDP simulations, the spin of the black hole is taken to
be $a/M=0.9$. The black hole is taken to be an AGN-class black hole with 
a mass of $10^8\,M_\odot$. Two broad classes of simulations were carried out,  distinguished by the stiffness of the equation of state; the Rad1 group
has a relativistic adiabatic index, $\Gamma=4/3$, while the Rad2 group has $\Gamma=5/3$, as was done in DHK03. 

\begin{table}[ht]
\caption{\label{simparams}Simulation Parameters.}
\begin{center}
\begin{tabular}{l|c|r|c|cccc}
\hline
Simulation & $a/M_{bh}$ & ${M_{bh}}$ & $\Gamma$ & Grid ($n_r\times n_\theta$)& $r_{in}/M_{bh} $&$r_{out}/M_{bh} $ &$\Delta \theta$\\
\hline
\hline
Rad1 & 0.900& $10^8\, M_\odot$    & 4/3 & $192 \times 192$ & 1.51 & 60.0 & $0.001 \pi$\\
Rad2 & 0.900& $10^8\, M_\odot$     & 5/3& $192 \times 192$ & 1.51 & 60.0& $0.001 \pi$\\
\hline
\end{tabular}
\end{center}
\end{table}

The key parameters for the initial tori are summarized in Table \ref{params}.
The initial state in each simulation is taken to be a 
$100\,M_\odot$ torus orbiting an AGN-like Kerr black hole. 
Each torus is initialized with a weak MRI (Magneto-Rotational Instability) seed field as measured by the
grid-averaged plasma $\beta$ of 100. Each torus has an inner edge, $r_{in}$, at $6.80\,M_{bh}$ (roughly
$3\,r_{ms}$, where $r_{ms} = 2.32\,M_{bh}$ is the marginally stable orbit of a Kerr black hole with spin $a/M_{bh}=0.9$).  The pressure
maximum in each simulation is at $12\,M_{bh}$, corresponding to an
orbital period of $276\,M_{bh}$, where $M_{bh}$ is the mass of the central
black hole. This choice of pressure maximum was made to hasten the growth of the MRI, and lies much closer to the black hole than in the analogous KDP simulation of DHK03. The density at the pressure maximum is on the order of $10^{-6}\,g\,cm^{-3}$.
The peak temperature of the initial torus is taken to be 
either $200$ K, $300$ K, $400$ K, or $1000$ K. These initial tori are labelled
somewhat arbitrarily cold (C), warm (W), hot (H), and very hot (V). The particular choice of temperatures was made to allow for the run of temperatures in the accretion flow to lie in the range of $1000 \lesssim T_{acc} \lesssim 10,000$ K, which corresponds with the temperature range of the zero-metallicity Rosseland mean opacity obtained by Lenzuni, Chernoff, \& Salpeter (1991; hereafter LCS91). The
typical densities in the initial tori also lie in the density range of this opacity function. For completeness, a reference
simulation (labelled R) is also carried out with radiative effects switched off; this 
corresponds to simulations as would have been carried out with the original DH03
GRMHD code. Each simulation is allowed to evolve through the saturation phase of MRI, which occurs after a few orbits of the torus/disk at the initial pressure maximum; in axisymmetric simulations the MRI is not sustainable past a few orbits, so the simulations where terminated after 
five orbits at the pressure maximum.

\begin{table}[ht]
\caption{\label{params}Initial Torus Parameters.}
\begin{center}
\begin{tabular}{l|c|r|r|c|c|c|c}
\hline
Model & Label & ${\rm T}_{disk}$& $M_{disk}$ & $\beta$ & $r_{in}/M_{bh}$ & ${r_P}_{max}/M_{bh}$ & 
$T_{orb}/M_{bh}$  \\
\hline
\hline
Ref   & R &$200\,K$ & $10^2\, M_\odot$& 100  & 6.80 & 12.0 & 276  \\
Cold  & C &$200\,K$ & $10^2\, M_\odot$& 100  & 6.80 & 12.0 & 276  \\
Warm  & W &$300\,K$& $10^2\, M_\odot$& 100 & 6.80 & 12.0 & 276   \\
Hot   & H & $400\,K$ & $10^2\, M_\odot$& 100 & 6.80 & 12.0 & 276 \\
Very Hot & V  & $1,000\,K$ & $10^2\, M_\odot$& 100 & 6.80 & 12.0 & 276 \\
\hline
\end{tabular}
\end{center}
\end{table}

For all simulations, the inner radial boundary is set at $1.51\,M_{bh}$, the outer radial boundary is at $60\,M_{bh}$, the polar boundaries are at 
$\pi \times 10^{-3}$ radians from the polar axes. Radial grid zones use exponential scaling; polar zones use a linear scaling; these have the effect of concentrating the computational
zones near the event horizon.
Outflow boundary conditions are in effect at
the inner and outer radial boundaries (the outer boundary imposes
an initial dust inflow until the outward accretion flow from the funnel/jet and
the outward motion of the outer edge of the accretion disk are established).
Reflecting boundary conditions are applied at the edges of the polar grid.
Periodicity is applied on the azimuthal grid faces.

\section{Results}

The simulations discussed in this paper are used to address a narrowly focused question:
what is the influence of radiation on mass flux in the accretion disk? Of
particular interest is the effect of radiation in the hot, dense plunging flow near the black hole. 

\subsection{Overall Evolution}

The set of ten simulations discussed here fall into two broad categories, sorted by
adiabatic index (see Table \ref{simparams}). Within each category, one
non-radiative and four radiative simulations
were carried out. The four radiative simulations are distinguished by the initial temperature of the torus.
As each simulation evolves, the overall characteristics described previously in DHK03
still apply (see, e.g., Figs. 2 and 3 of DHK03): the seed magnetic field in the initial torus is 
amplified by shearing; during
the first orbit, a thin stream of accreting matter threads its way along the equatorial
plane towards the black hole. The Magneto-Rotational Instability (MRI; Balbus \& Hawley, 1998)
peaks during the second orbit of the main disk, driving a very active turbulent phase where
angular momentum is transported outwards. With peak MRI activity, a good deal of
matter is driven towards the black hole, and at the same time the outer edge of the initial
torus begins an outward migration. In addition, a two-component outflow develops in the evacuated funnel,
consisting of a relativistic, low density component along the axes of the black hole, and
a denser, slower outflow along the funnel wall. Since the MRI is not sustainable in 
axisymmetry, turbulence subsides after a few orbits of the main disk. Peak activity
occurs between two and three orbital periods at the initial pressure maximum.

As gas spirals on along the equatorial plane towards the
black hole, it is compressed and heated, with temperature rising by an order of magnitude or so
as the flow converges on the inner radial boundary. Since the radiation tensor in the
diffusion approximation is set by the temperature and density of the gas, radiative effects ``follow'' 
the accretion flow and build up in the plunging region because
of the elevated temperature there. However, radiative effects are also quite noticeable within the body of
the disk as well, especially near the expanding outer edge where density drops and
temperature rises.

\subsection{Density Profile\label{Density}}

\begin{figure}[htbp]
\begin{center}
\includegraphics[width=3.in]{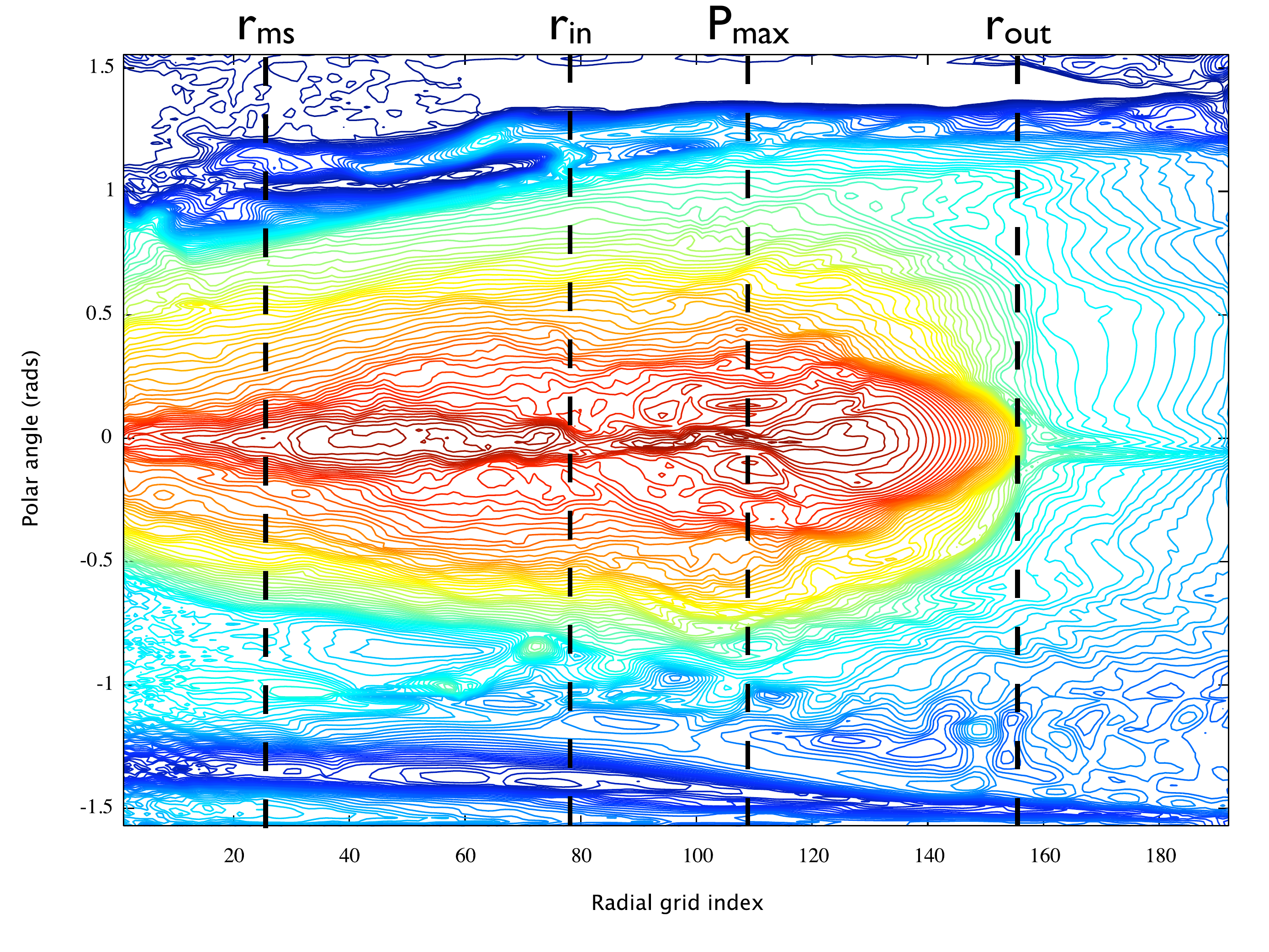}
\includegraphics[width=3.in]{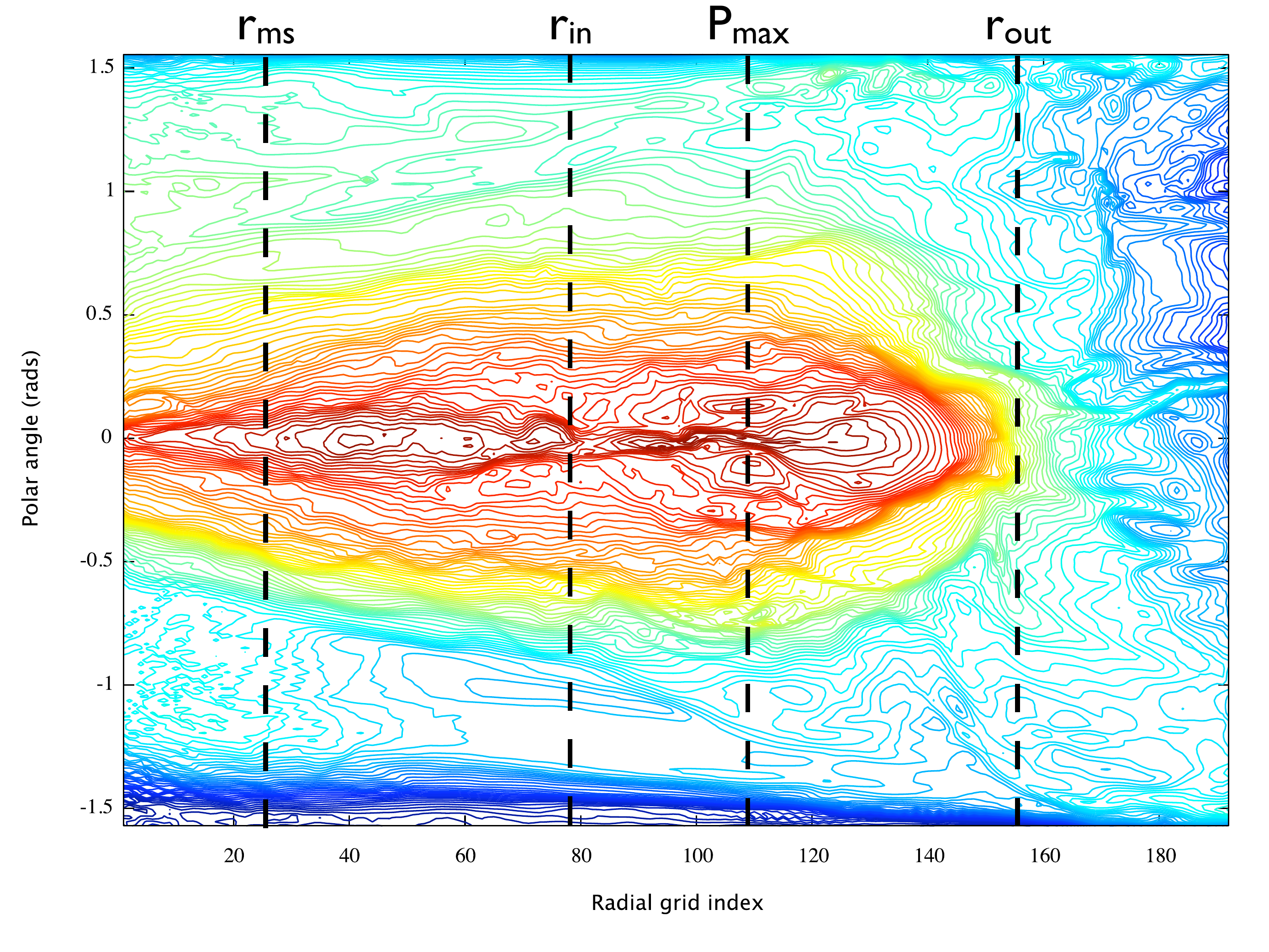}
\caption{Density profiles, $\overline{\rho}(r,\theta)$, averaged over one orbit, comparing the reference run
Rad1R (left panel) and the hot simulation Rad1H (right panel). The average is taken between $t=2\,T_{\rm orb}$ to $3\,T_{\rm orb}$ as measured at the initial
pressure maximum. Contours are logarithmic and are equally spaced over 10 decades, scaled to the initial maximum density. Plots are shown in coordinate space, with the polar angle referenced to the equatorial plane. For reference, vertical dashed lines indicate the location of the marginally stable orbit, as well as the inner edge,
pressure maximum and outer edge of the initial torus.}
\label{DensityAvg}
\end{center}
\end{figure}

The presence of radiation alters the distribution of gas within the disk. This can be seen in Figure~\ref{DensityAvg}, where plots of density are shown
for two Rad1 simulations, reference simulation Rad1R and its hot counterpart
Rad1H. The plots are time-averaged over the interval from
$t=2\,T_{\rm orb}$ to $3\,T_{\rm orb}$, corresponding to peak MRI activity. 
The plots are shown in coordinate space, which emphasizes the region inward
of the pressure maximum, where the radial grid is concentrated. 

Though the overall structure of the
denser parts of the accretion disks is similar, as shown by the family of red contours, it is apparent that the funnel contains
denser gas in Rad1H than it does in Rad1R, so even though radiative effects are switched off in the unbound funnel, differences due to radiation in the bound portion can influence the
structure of the funnel (this underscores the importance of a proper radiative treatment
in all regions of the simulation volume). 
In addition, there is a notable difference in the structure of the disk at large radii. It appears as if the radiative simulation
shows a disk with a greater outward transport of material. Any differences in the structure
of the plunging flow are hard to discern in these plots.

Similar outcomes, altered funnel density and
greater outward migration of the disk at large radii, are seen in all radiative simulations. 

\subsection{Temperature Profile\label{Temp}}

\begin{figure}[htbp]
\begin{center}
\includegraphics[width=3.in]{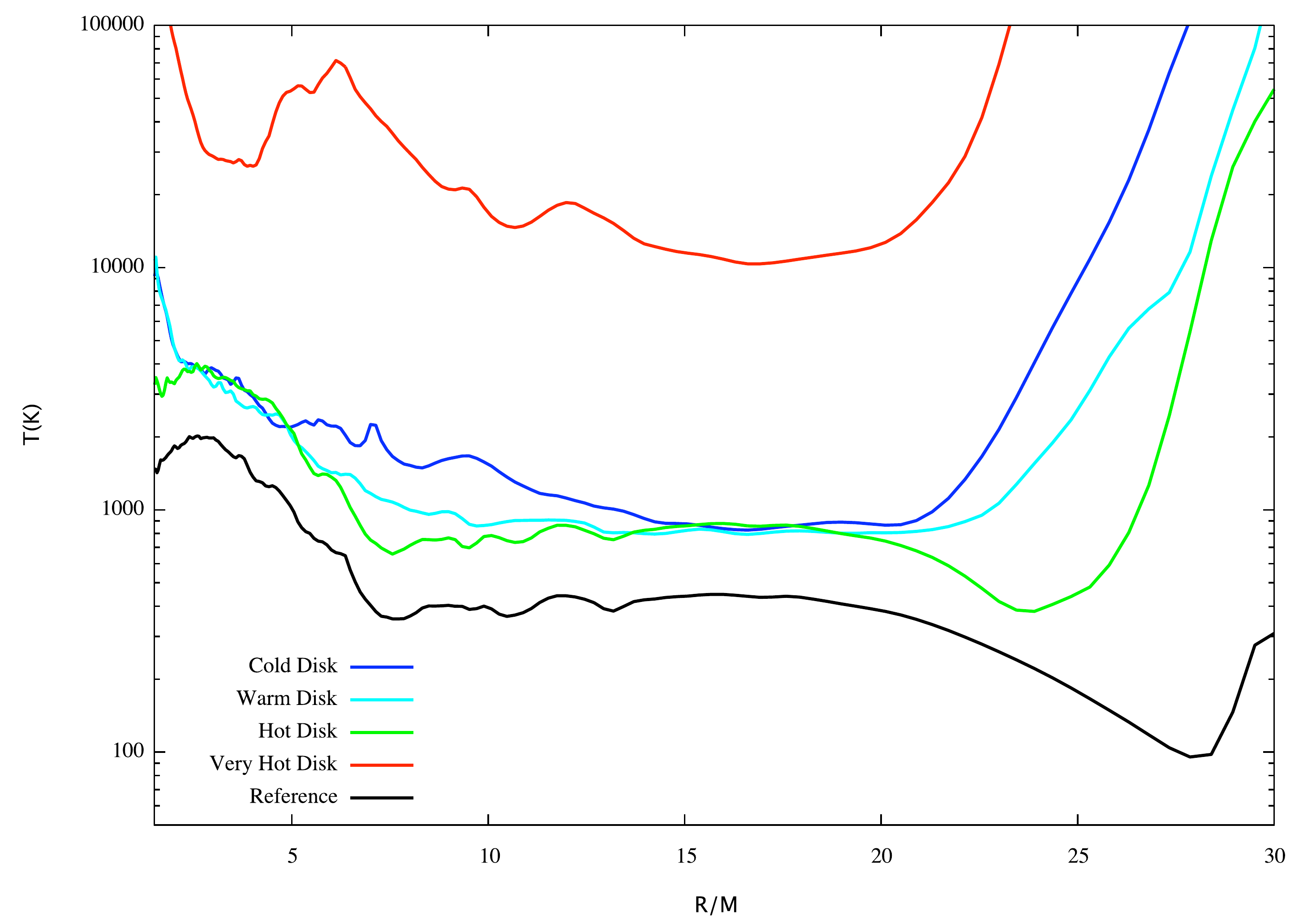}
\includegraphics[width=3.in]{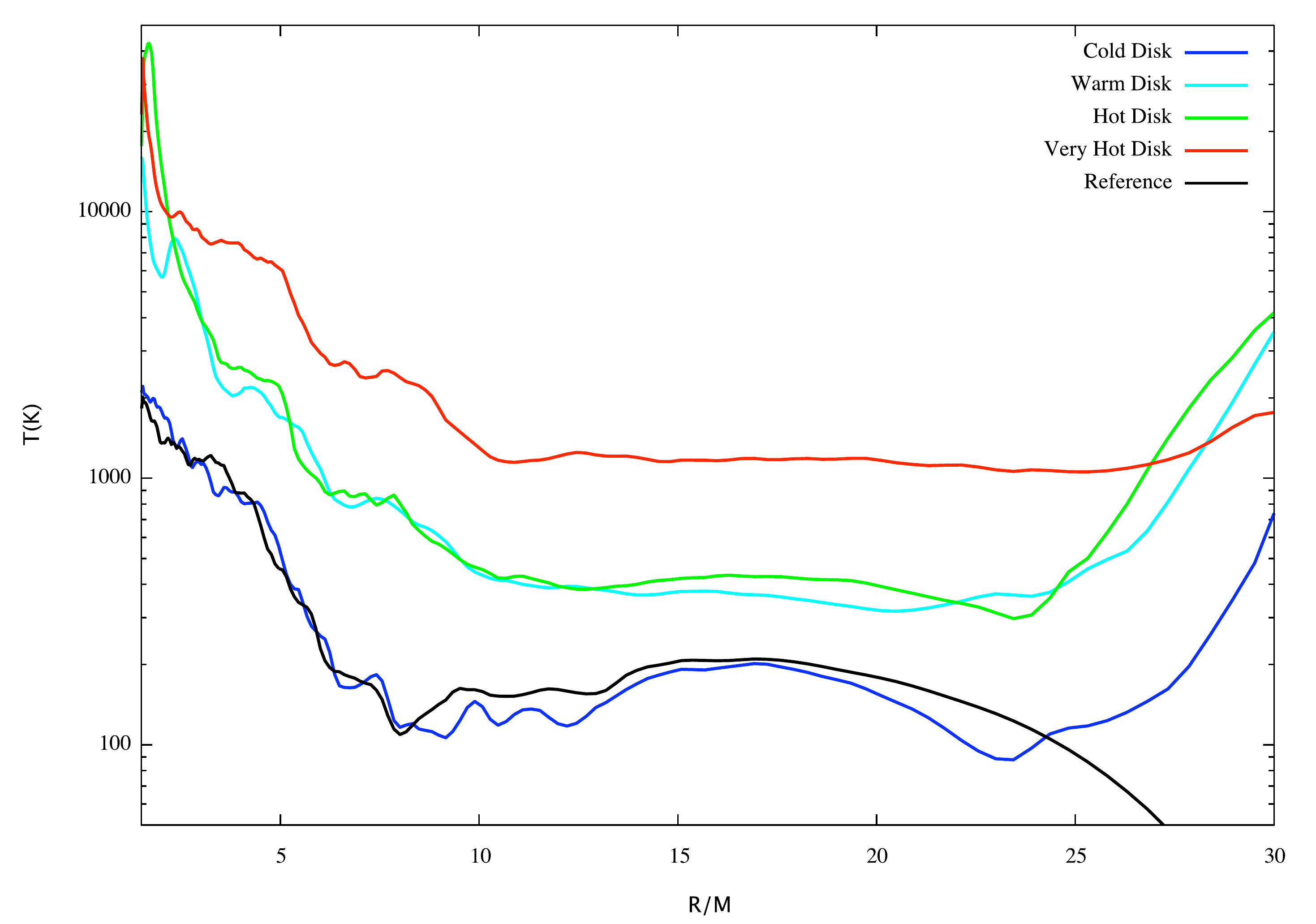}
\caption{Run of temperature near equator for Rad1 (left panel) and Rad2 (right panel) simulations. The reference simulation in each family is plotted using the 
same temperature scaling factor as the corresponding cold simulation.}
\label{Temperature}
\end{center}
\end{figure}

The radial run of temperature covering the plunging flow and the main disk body is shown in Figure~\ref{Temperature} for all
simulations. Each curve represents averages of temperature in the
vicinity of the equatorial plane. Steep rises in temperature are seen both in the
direction of the black hole, and also in the bound, low density outer region of the disk (the outer edge of the initial torus lies at $r=30\,M_{bh}$). 
Since radiative effects in the diffusion approximation are directly tied to local temperature,
it is to be expected that both the plunging flow and the outer edge of the disk should be
locales where radiative effects might show the greatest divergence from the
reference simulations. The reference simulations are plotted using the same
temperature scaling as their cold counterparts (i.e. assuming a
$200\,K$ non-radiating initial torus).
 
The temperature distribution in each simulation shows unique characteristics
due to the interaction of the radiative terms with the other components of the
GRMHD equations; the curves are not merely shifted vertically by the different temperature
scalings used in each simulation. Though the temperature profile through the
main disk body ($10\,M_{bh} < r < 20\,M_{bh}$) is relatively flat in all
simulations,
there is an order of magnitude rise from the inner edge of the disk to the
inner radial boundary in the Rad1 simulations, and a steeper rise in the
Rad 2 simulations. There is also a steep rise in temperature
from $r \approx 25\,M_{bh}$ outward; the rise is steeper in the Rad1 family than
in the Rad2 family. In both families of simulations, this upturn in temperature at large radii is unique to the 
radiative simulations, and is absent from the reference simulations. In the Rad2
family, the Rad2R and Rad2C temperature profiles closely parallel one another
for $r < 25\,M_{bh}$, while in the Rad1 family, the Rad1R and Rad1C profiles are quite distinct.

\subsection{Accretion Rate}

\begin{figure}[htbp]
\begin{center}
\includegraphics[width=3.2in]{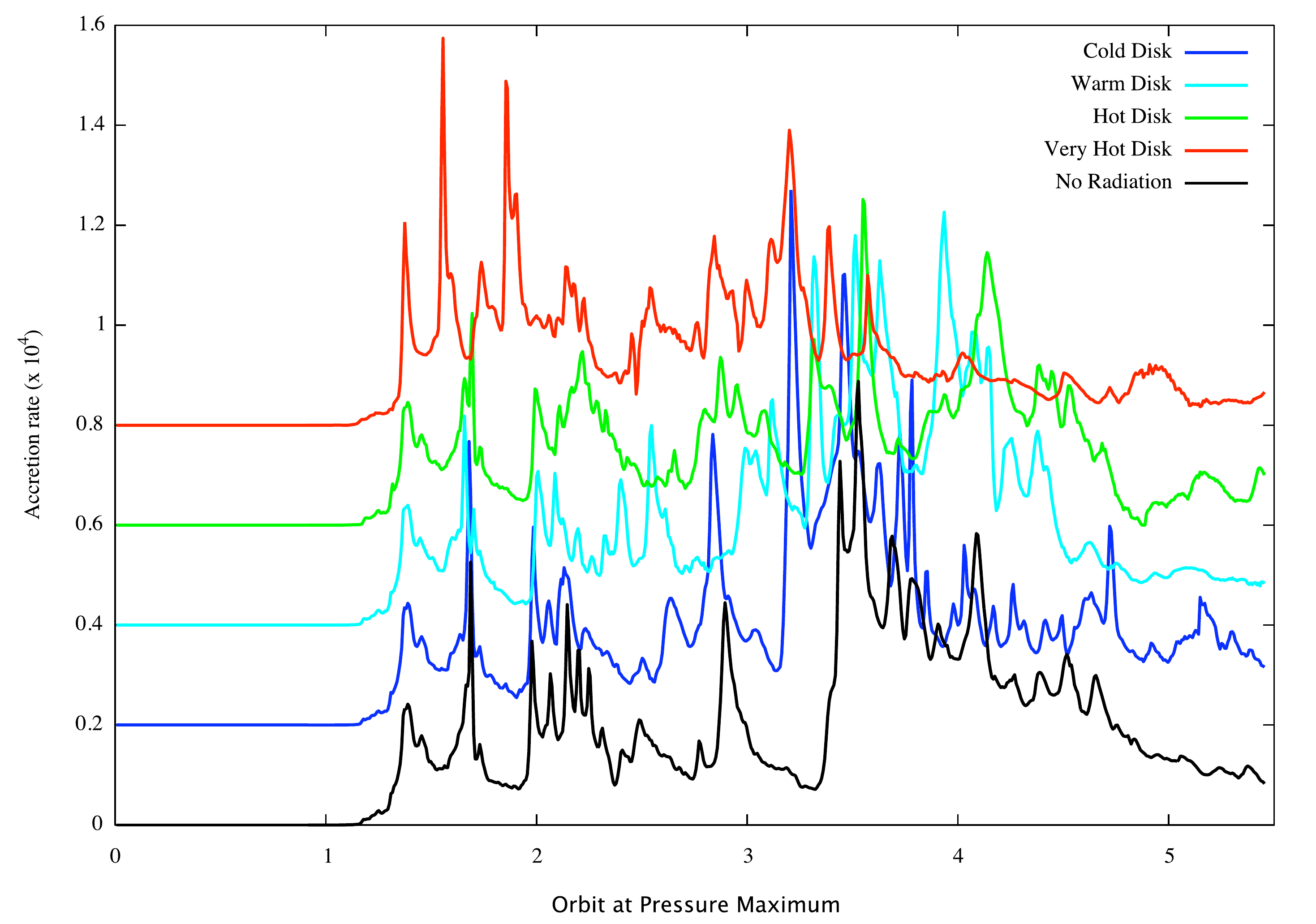}
\includegraphics[width=3.2in]{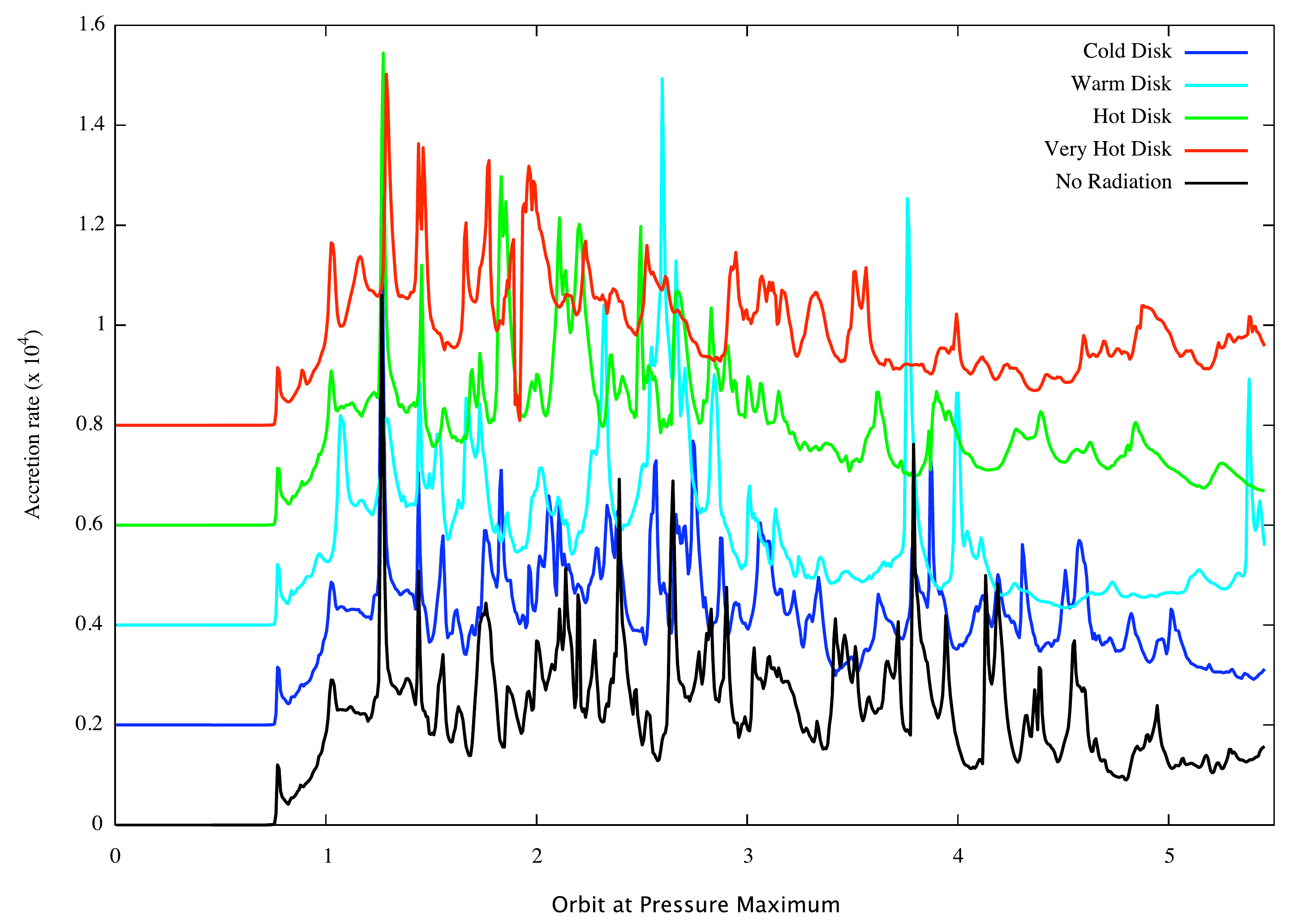}
\caption{Accretion rate $\langle -\rho\,U^r\rangle(r_{inner},t)$ through inner radial boundary for Rad1 (left panel) and Rad2 (right panel) simulations. Accretion rate is expressed as a fraction of the initial torus mass.}
\label{DDot}
\end{center}
\end{figure}

Perhaps the simplest diagnostic to quantify the role of radiation in the
simulations is the shell-averaged radial transport of matter, $\dot{M}=\langle \rho\,U^r\rangle(r,t)$.
This diagnostic, when plotted near the inner boundary, measures the rate
of accretion onto the black hole as a function of time. Figure~\ref{DDot} shows the 
accretion rates for the Rad1 and Rad2 families. Though the graphs show 
a good deal of variability, it is nonetheless possible to see that notable
departures from the reference simulations take place when radiative effects are
added.
The accretion rate is zero during the first orbit of the Rad1 tori while the
leading edge of the plunging flow makes it way towards the black hole. Material arrives slighly earlier for the Rad2 simulations. The rate of accretion is
highly variable in all simulations, with several sharp peaks marking the
arrival of dense knots of material from the accretion disk. In the early
stages (between orbits 1 and 2), all accretion rate curves are essenitally identical; as the plunging flow builds up and temperature rises, radiative effects begin to assert themselves. Beyond orbit 2, there are notable differences in the number, location, and amplitude of peaks in accretion rate
between the simulations, though any trends are difficult to make out in these
plots. In all cases, the rate of accretion decreases both in intensity and variability after four orbital
periods due to a decrease in MRI-driven turbulence.  The peak accretion rates, with values on the order
of $10^{-4}$, agree with the full 3D simulations of DHK03 as well as axisymmetric simulations
of disks embedded in large-scale fields reported by De Villiers (2006).

A more quantitative measure is obtained by summing the $\langle -\rho\,U^r\rangle(r_{inner},t)$
diagnostic over time to obtain a measure of total accreted mass. These sums, normalized
to the reference simulations Rad1R and Rad2R are given in Table \ref{accretedmass}.
\begin{table}[ht]
\caption{\label{accretedmass}Relative Accreted Mass, ${M}_{\rm rad}/{M}_{\rm ref}$.}
\begin{center}
\begin{tabular}{c|c|c}
\hline
Model/        & Rad1  & Rad2\\
Simulation & ($\Gamma=4/3$)  & ($\Gamma=5/3$)\\
\hline
\hline
C  & 1.03 & 0.96  \\
W  & 1.08 & 0.74 \\
H   & 0.77  & 0.83 \\
V  & 0.61 & 0.77  \\
\hline
\end{tabular}
\end{center}
\end{table}
It is clear from the table that the cold simulations Rad1C and Rad2C do not significantly differ in
overall accretion from the reference simulations. However, important differences arise for the hotter initial
tori, which build up higher temperatures in the plunging region. The Rad1 models
($\Gamma=4/3$) seem to show a steady drop in accreted mass with higher initial temperature,
while the Rad2 models ($\Gamma=5/3$) show a reduced amount of accreted mass for the W, H, and V simulations,
without an apparent trend.

\begin{figure}[htbp]
\begin{center}
\includegraphics[width=3.in]{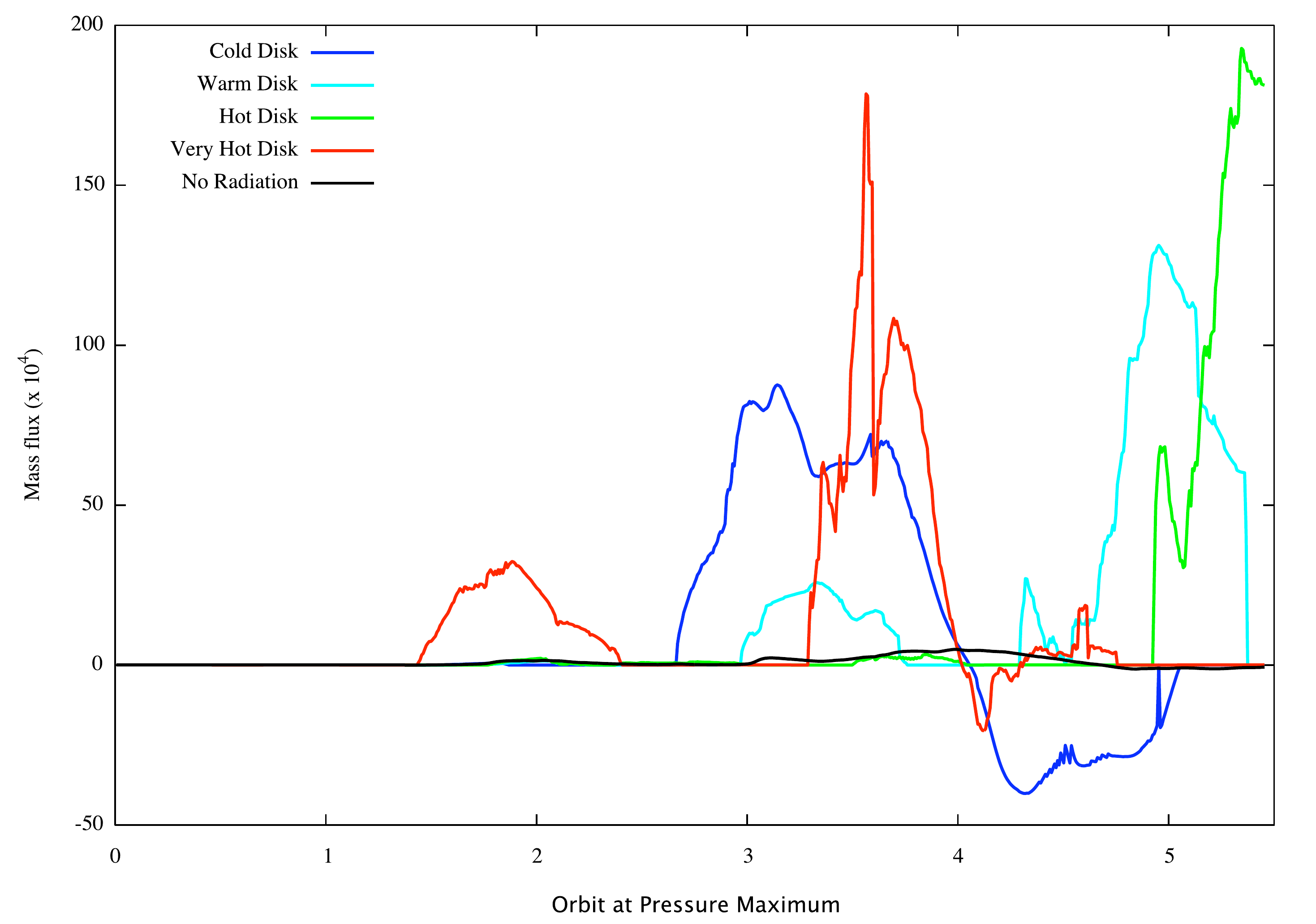}
\includegraphics[width=3.in]{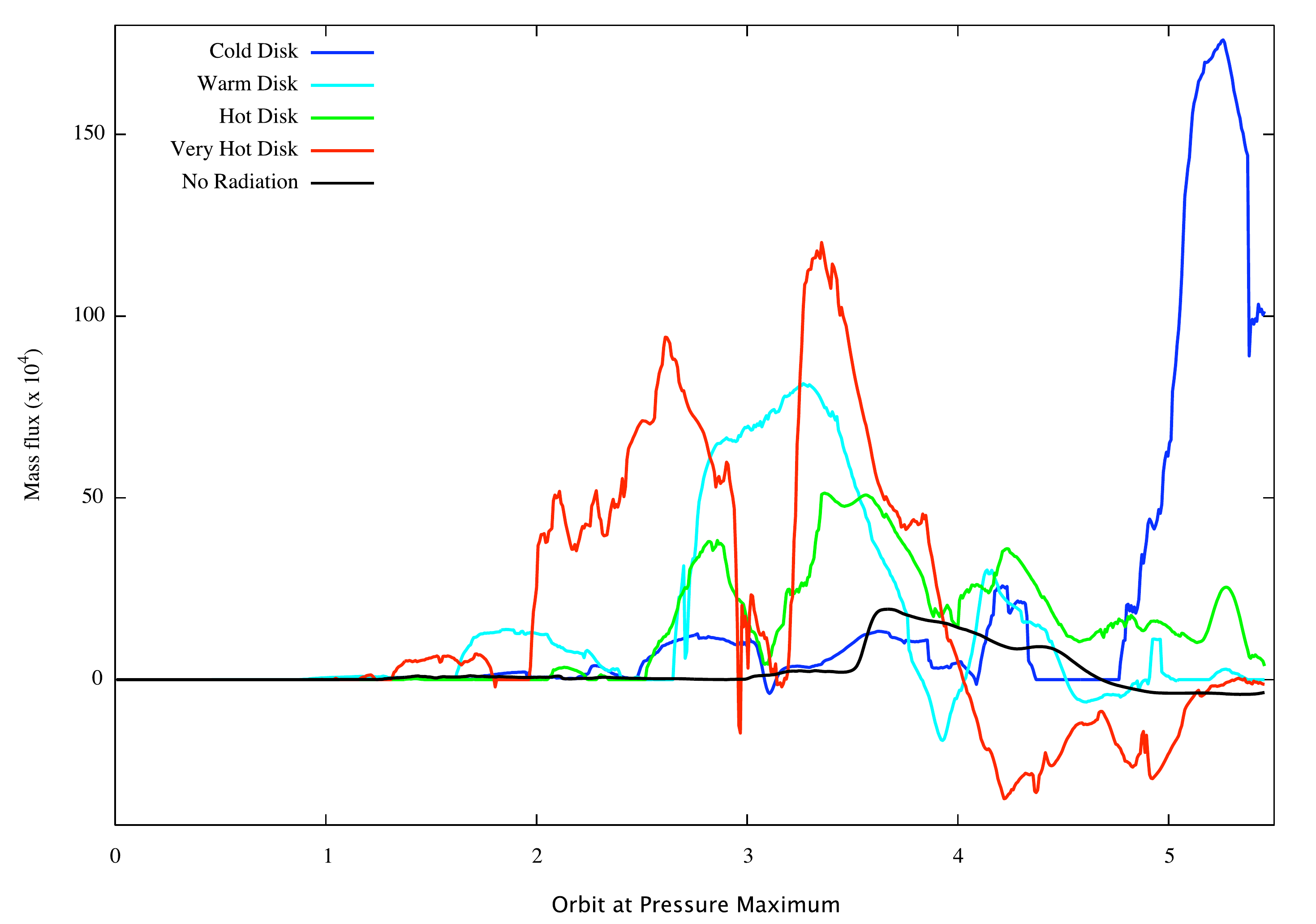}
\caption{Mass flux $\langle \rho\,U^r\rangle(30\,M_{bh},t)$ past outer edge of initial torus for Rad1 (left panel) and Rad2 (right panel) simulations. Mass flux is expressed as a fraction of the initial torus mass.}
\label{DDotLargeR}
\end{center}
\end{figure}

As noted in \S\ref{Density}, the density profiles suggest that a greater density is found at
large radii in the radiative simulations than in the reference simulations. Also, as noted in \S\ref{Temp}, a steep rise in temperature is also seen in
the outer portions of the disk, suggesting that radiation is acting to increase
the temperature in this region. Are the two related? Figure \ref{DDotLargeR} 
shows the mass flux diagnostic plotted at $r=30\,M_{\rm bh}$; this tracks the passage of matter at a point just outside the outer edge of the initial torus for the Rad1 and Rad2 families of simulations. 
The departure from the reference simulation (black line) is remarkable. Even though there are large excursions in both directions, radiative
effects in this region significantly enhance the net outward flow of matter from within
the core of the accretion disk.

\subsection{Radiation Field}

\begin{figure}[htbp]
\begin{center}
\includegraphics[width=3.2in]{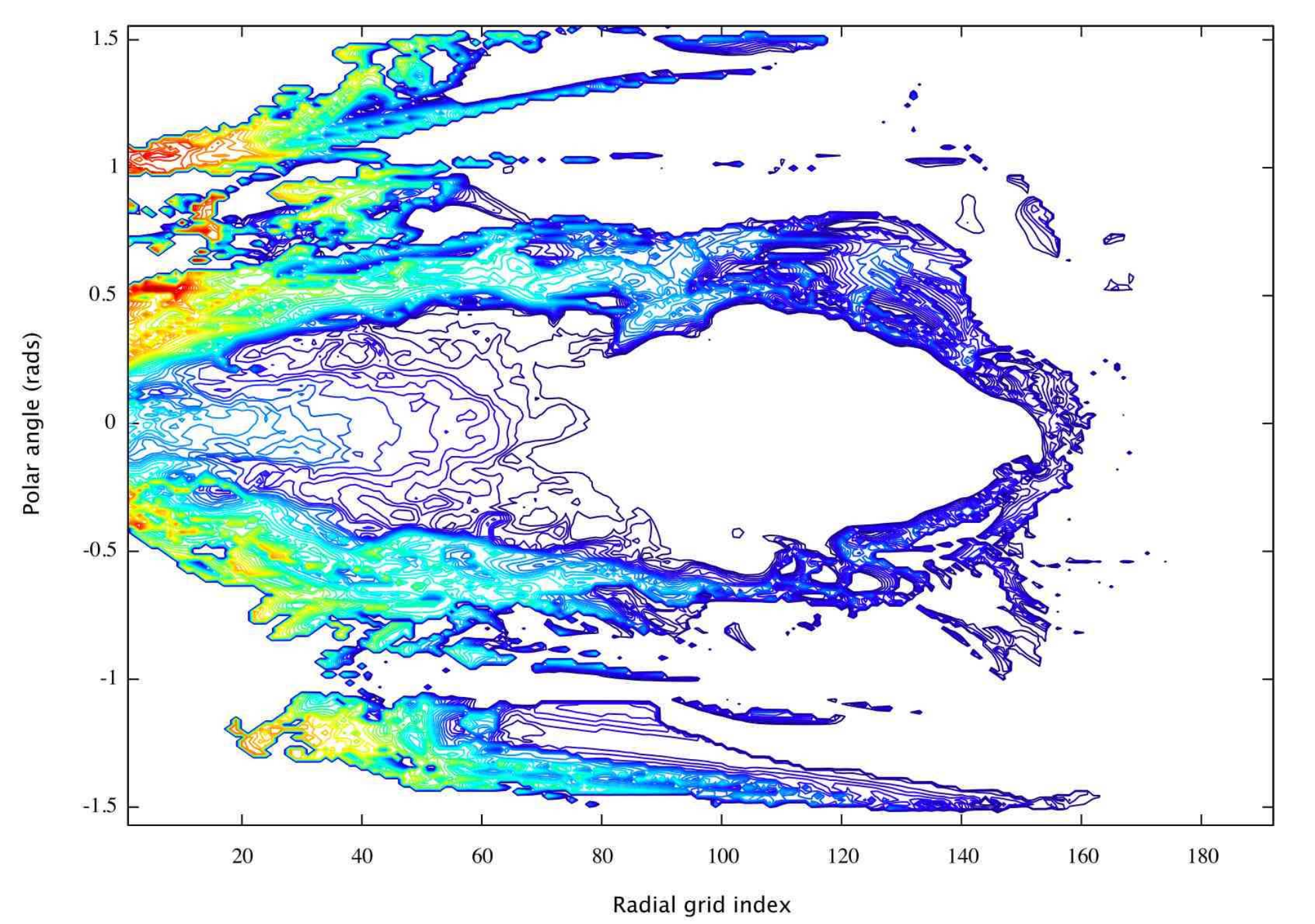}
\includegraphics[width=3.2in]{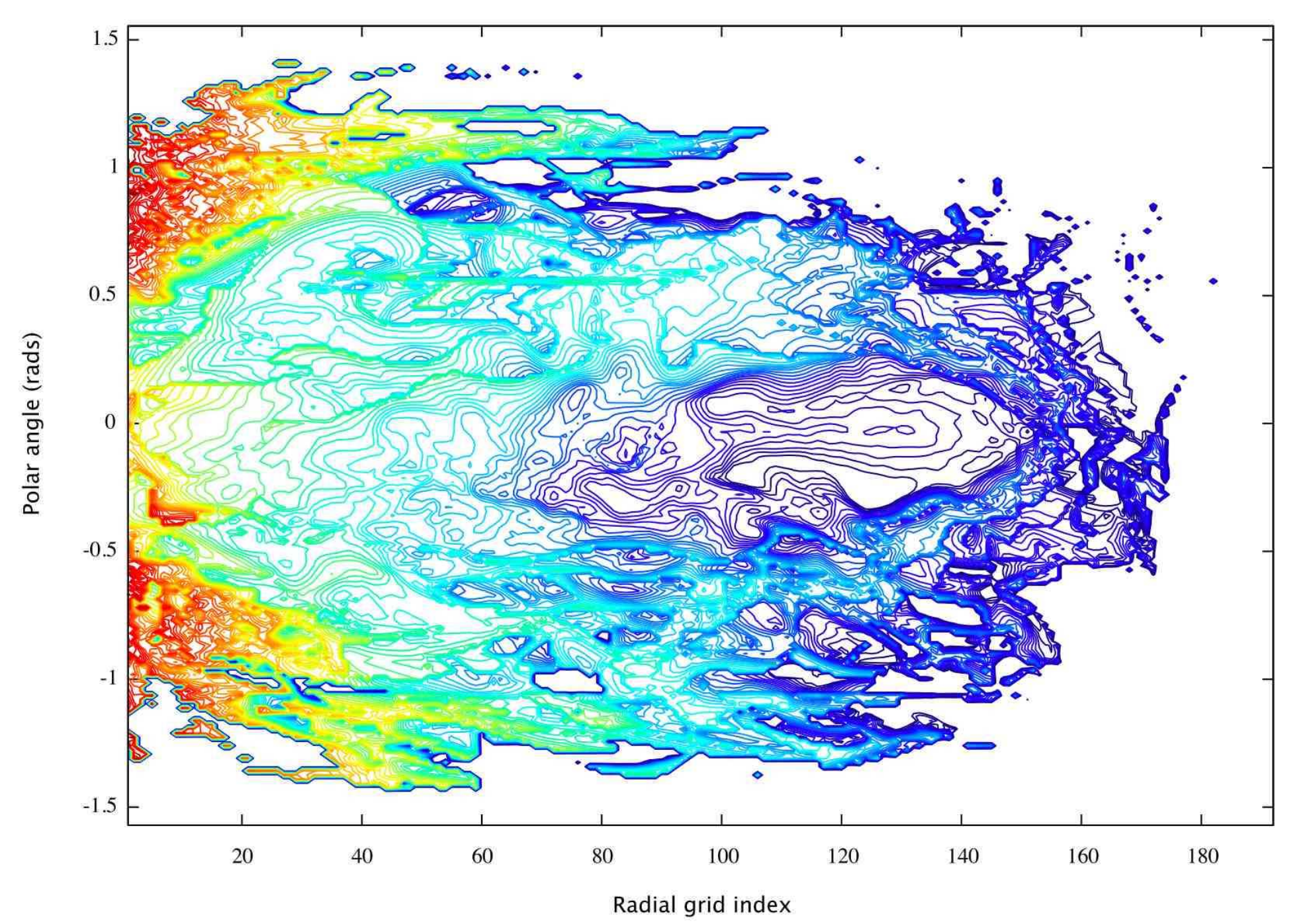}
\caption{Plot of time-averaged $\mathbf{R}^{tt}(r,\theta)$ for the Rad1H (left panel) and 
Rad2H (right panel) simulations. The average is taken between $t=2\,T_{\rm orb}$ to $3\,T_{\rm orb}$ as measured at the initial
pressure maximum. Contours are logarithmic and are equally spaced over 13 decades. Plots are shown in coordinate space, with the polar angle referenced to the equatorial plane. Both plots use the same absolute log scaling. }
\label{E0Avg}
\end{center}
\end{figure}

All components of the radiation tensor are dumped periodically, along with
other code variables. Figure \ref{E0Avg} shows the time-averaged structure
of the radiative energy density (the time-time component of the radiation tensor, $\mathbf{R}^{tt}$), in the Boyer-Linquist frame for the Rad1H and Rad2H simulations; similar plots are obtained for the other simulations. The figure shows the effect of the switch controlling the radiative effects: the white patches at high and low polar angles are unbound regions where the radiation tensor is zeroed out. In both plots, the largest values of the radiative energy density straddle the plunging flow, in a region where the density is lower and temperature hotter than in the plunging flow. There is also a visible increase in intensity along the equator towards the black hole. Large streaks of bound material also occupy the funnel region in the two simulations; they represent the passage of clumps of bound, hot gas through the funnel (clumps of bound gas have turning points at large radii, but for these simulations, they exit the outer radial boundary). 

It is apparent from this figure that the boundary between the bound and unbound portions of the simulation volume could give rise to very large gradients in the components of the radiation tensor, and could prove destabilizing. To guard against this, the portions of the code implementing radiative updates in the energy and momentum equations detect and prevent steep gradients due to zero values in the radiation tensor from entering the numerical solution. This limiter will be removed when
the full version of the rGRMHD code is implemented.

\section{Discussion and Conclusions}

This brief paper has introduced a set of simulations meant to perform an
initial validation of a new rGRMHD code that incorporates radiative effects into the dynamical equations, but does not yet implement the numerical solution of the Radiative Transfer Equation (RTE). By using the diffusion approximation, it is possible to prescribe the radiation tensor based on local properties of the fluid, without requiring a solution of the RTE. Even in this partial implementation, the rGRMHD code tracks properties of the radiation field
(energy density, pressure, fluxes), and reports these through the history diagnostics and binary dumps. Quantitative analysis of in-situ effects of radiation is now possible, and with the ray tracing capabilities built into the code, detailed simulated observations are also within reach (though, as discussed in the appendices, such simulated observations are much more meaningful in full 3D simulations with sizable simulation volumes). 

Though the numerical solutions produced a large amount of diagnostic information, this preliminary analysis focused on a very simple question: does the presence of radiation affect the transport of matter through the accretion disk? 

Qualitatively, the simulations have shown that density and temperature in the accretion disk and plunging region are affected by the radiation field: details of
the run of temperature in the plunging region and in the outer region of the disk depend on the initial strength of the radiation field (which is set by the model-dependent disk temperature); radiation is also found to enhance the outward transport of matter from the accretion disk. The cold initial tori tend to track the evolution of the reference simulations (which do not include
radiative effects), suggesting the the radiative results converge on the 
non-radiative results for weak radiation fields. The agreement between 
cold and reference simulations is best for the simulations using $\Gamma=5/3$.
Significant departures in density and temperature profiles are found for
stronger initial radiation fields (i.e. hotter initial tori).

Quantitative analysis relied on a well-exercised
diagnostic, radial mass flux $\langle\rho\,U^r\rangle(r,t)$, as a means of comparing radiative and non-radiative simulations.
This diagnostic shows that the amount of matter accreted onto the black hole and the amount of matter transported outward from the main body of the disk are affected by radiation: accretion near the event horizon is reduced and matter transport at large radii is enhanced. For the simulations using a relativistic equation of state ($\Gamma=4/3$) there seems to be a reduction in accretion that scales with temperature (hotter plunging flows tend to produce less accretion onto the black hole); for the adiabatic equation of state ($\Gamma=5/3$), accretion is reduced by radiation from a hot inflowing gas, but the evidence for a clear temperature trend is not obvious. 
The accretion rates shown in Figure \ref{DDot}, when summed over the orbit where peak MRI activity occurs (between $t=2\,T_{\rm orb}$ to $3\,T_{\rm orb}$ as measured at the initial pressure maximum), correspond to a rate on the order of one solar mass per year, a number that seems reasonable for relatively cold disks feeding AGN-scale systems. 

It would be interesting at some later time to investigate whether
accretion can be completely shut off by radiative effects in these
types of simulations, and to investigate this outcome in relation to
the general analyses of the Eddington limit (Frank, King, \& Raine, 2003).
This would most likely require the full rGRMHD code (with RTE), or at least
the current code with an opacity function spanning a greater temperature range
than LCS91. 
Ultimately, it is hoped that a proper treatment of the RTE will help shed light on the role of radiation not only in regulating accretion, but also in powering the jets and funnel outflows; though these features have already been noted in the non-radiative DHK simulations, it remains important to clarify the role of radiation in these energetic processes. Figure \ref{E0Avg} is especially
suggestive in this regard: though the effects described here apply to the
dense plunging flow, the radiation field is very intense in the region straddling this flow (and would likely be even greater at the base of the
funnel if the current implementation did not switch off the radiation field there),
so that the jets emerging from a fully radiative treatment would likely
receive a substantial push from the intense radiation above the black hole
pole caps.

The initial tori chosen for these simulations are very compact and produce
prompt accretion. This was done by design to achieve a rapid turn-around time, and is justified because of the emphasis on accretion rate in the plunging flow. Drawing conclusions beyond this narrowly framed question seems
inappropriate.
Caution is also in order even in this limited interpretation of the
simulations. The original GRMHD code has grown substantially in
complexity, and a large number of diagnostics have been added, so that it will
take time to develop a body of knowledge around this new data. For the
time being, it seems wise to err on the side of caution when interpreting simulation outcomes by looking primarily for effects that reveal unambiguous trends across families of simulations when referenced to the existing base
of knowledge from non-radiative GRMHD simulations.

As a final note, in recent years, a number of studies have relied on the accretion rate produced by non-radiative simulations as a proxy for dissipation, or have used the outcome of non-radiative simulations and applied ray tracing
methods is a post-processing phase to approximate a radiating fluid. Figures~\ref{DensityAvg}, \ref{Temperature}, and \ref{DDot} show that radiation alters the detailed structure and time-dependence of the accretion flow from the non-radiative reference simulations. If nothing else, these simulations have demonstrated that reliance on post-processing of non-radiative simulations for emission studies is of limited use: radiation can substantially alter an accretion flow and must be treated in a self-consistent manner.

{\bf Acknowledgements:} 

Although the development of the full rGRMHD code is an independent research project, 
preliminary work on the
ray tracing component was undertaken while the
author was at the University of Virginia (2000-2004), supported by
NSF grants AST-0070979 and PHY-0205155, and NASA grant NAG5-9266 (Principal
Investigators John Hawley and Steve Balbus). 

All simulations were carried out on an
8-core Mac Pro 3 GHz Xeon system with 2GB of RAM. Source code for this project was compiled with the GNU gfortran compiler and OpenMPI libraries. Graphics 
for this paper were generated using the GNU Octave application.

\newpage
\appendix{{\Large{\bf Appendices - Building The rGRMHD Code}}}

Introducing a radiation mechanism to the GRMHD code is not
a trivial undertaking and the process has been broken down into a
series of steps, each outlined in the following appendices. This paper
builds upon the discussion in DH03 and DHK03, and the reader unfamiliar with the
original GRMHD code should refer to these papers prior to reading the
following descriptions. A general derivation of the equations of radiative hydrodynamics 
in the Kerr metric can be found in T07, and complements 
the numerically oriented treatment discussed here in the MHD context.

\section{Equations of Radiative GRMHD\label{radequations}}

In the GRMHD code of DH03, the state of the relativistic test fluid at each point in the spacetime
is described by its density, $\rho$, internal energy, $\epsilon$,
$4$-velocity $U^\mu$, and isotropic pressure, $P$, which is related to
the first two scalars via the equation of state of an ideal gas,
$P=\rho\,\epsilon\,(\gamma-1)$, where $\gamma$ is the adiabatic
exponent. The relativistic enthalpy is $h=1 + \epsilon + P/\rho$. This set of variables
needs to be augmented to include the presence of a radiation field.

The rGRMHD code continues to use the Boyer-Lindquist coordinate system as the
main reference frame for the simulations. However, radiative effects are best calculated
in the frame of reference that is locally comoving with the radiating fluid (MM84), necessitating
the introduction of coordinate transformations that will be discussed in greater detail in
Appendix \ref{radtensor}. In the following description, all vector and tensor quantities are expressed
in the Boyer-Lindquist reference frame. The Boyer-Linquist frame constitutes a 
coordinate basis, for which the following identities for the divergence of a four-vector and a tensor can be used to simplify expressions:
\begin{equation}\label{divvec}
{\nabla}_{\mu}\left(f \,{v}^{\mu}\right) = {1 \over \sqrt{-g}}\,
\partial_{\mu}\left( \sqrt{-g} \,f \, {v}^{\mu}\right),
\end{equation}
\begin{equation}\label{divten}
{\nabla}_{\mu}\left({\beta}^{\mu\,\nu}\right) = {1 \over \sqrt{-g}}\,
\partial_{\mu}\left( \sqrt{-g}\,{\beta}^{\mu\,\nu}\right)+
{\Gamma}^{\nu}_{\epsilon\,\mu}\,{\beta}^{\mu\,\epsilon} .
\end{equation}
Greek indices range over all four spacetime coordinates, while roman
indices range over spatial coordinates only.

The equations of ideal GRMHD are the law of 
baryon conservation,
\begin{equation}\label{barcons}
{\nabla}_{\mu}\,(\rho\,U^{\mu}) = 0 ,
\end{equation}
where ${\nabla}_{\mu}$ is the covariant derivative, 
the conservation of stress-energy,
\begin{equation}\label{tmncons}
{\nabla}_{\mu}{T}^{\mu\,\nu} = 0 ,
\end{equation}
where ${T}^{\mu\,\nu}$ is the energy-momentum tensor for the fluid, 
and the induction equation,
\begin{eqnarray}\label{ct.3a}
\partial_j \left({\cal{B}}^j\right) & = 0 & (\nu=0) ,\\
\label{ct.3b}
\partial_t \left({\cal{B}}^i\right) -
\partial_j \left(V^i\,{\cal{B}}^j-{\cal{B}}^i\,V^j\right)& = 0 & (\nu=i),
\end{eqnarray}
where ${\cal{B}}^i$ are the components of the Constrained-Transport magnetic
field, and the transport 
velocity (also known as the coordinate velocity) $V^\mu$ is defined
$V^\mu = U^\mu / U^t$, where $U^t = W/\alpha$, and $W$ is the relativistic gamma-factor.
The magnetic field is also described by the
magnetic field $4$-vector, $b^\mu$.  The latter is fundamental to the
definition of the total four momentum,
\begin{equation}\label{momdef}
 S_\mu = (\rho\,h\ + {\|b\|}^2)\,W\,U_\mu ,
\end{equation}
and the  normalization condition
\begin{equation}\label{momnorm}
g^{\mu \nu}\,S_\mu\,S_\nu = -{\left[(\rho\,h+{\|b\|}^2)\,W\right]}^2 ,
\end{equation}
which is algebraically equivalent to the usual normalization
condition $U^\mu U_\mu = -1$. 
We define auxiliary density and energy functions $D =
\rho\,W$ and $E = D\,\epsilon$.  The set of variables $D$, $E$,
$S_\mu$, ${\cal{B}}^i$, $V^i$, and $b_\mu$ constitute the fundamental
GRMHD code variables.

The equations of GRMHD are augmented by the contribution from
a radiation field in a straightforward manner, by adding a radiative
contribution to the energy-momentum tensor,
\begin{equation}
\mathbf{T}^{\mu \nu} = \mathbf{T}_{\rm (fluid)}^{\mu \nu} + \mathbf{T}_{\rm (EM)}^{\mu \nu} + \mathbf{R}^{\mu \nu}
\end{equation}
where 
\begin{eqnarray} \label{tmndef}
\mathbf{T}_{\rm (fluid)}^{\mu \nu}& = & \rho\,h\,{U}^{\mu}\,{U}^{\nu}+ P\,{g}^{\mu\,\nu}\\
\mathbf{T}_{\rm (EM)}^{\mu \nu} &= & \left(
{1 \over 2}\,g^{\mu \nu}\,{\|b\|}^2+U^\mu\,U^\nu\,{\|b\|}^2-
 b^\mu\,b^\nu \right)\\
\mathbf{R}^{\mu \nu} &= &
\left[
\begin{array}{c|c}
  \mathcal{E}& \mathcal{F}^j \\
\hline 
  \mathcal{F}^i& \mathcal{P}^{ij}   
\end{array}
\right]
\end{eqnarray}
where the radiation tensor, $\mathbf{R}^{\mu \nu}$, contains $\mathcal{E}$ is the radiative energy
density, $\mathcal{F}$ the radiative flux, and  $\mathcal{P}$ the radiative stress;
respectively the zeroth, first, and second moments of the radiation field. 
These components are evaluated using
\begin{equation}
\mathbf{R}^{\mu \nu} =\int \oint \mathcal{I}(\mathbf{n},\nu)\,n^\mu\,n^\nu\,d \Omega\,d \nu
\end{equation}
where $\mathcal{I}$ is the intensity of the radiation field, $\mathbf{n}$ is the direction 4-vector. In a complete treatment
of radiation, the intensity is given by the radiative transfer equation (RTE).
In the current implementation, evaluation of the radiation tensor is greatly
simplified by using the first-order diffusion approximation,
which directly prescribes $\mathbf{R}^{\mu \nu}$ from the temperature and
mean opacity of the gas  (see Appendix \ref{diffusion}). Inclusion of the RTE is
the subject of ongoing work and is not described here.

Radiative contributions enter the GRMHD code through the energy and 
momentum equations, and the derivation proceeds as described in DH03.
The equation of energy conservation if obtained by projecting the conservation
law $\nabla_\mu \mathbf{T}^{\mu \nu} = 0$ onto the fluid 4-velocity:
\begin{equation}
U_\nu\nabla_\mu \mathbf{T}^{\mu \nu} = 0,
\end{equation}
which expands to
\begin{equation}\label{eq.1a}
{U}_{\nu}\,{\nabla}_{\mu}{\mathbf{T}}^{\mu\,\nu} =
 {U}_{\nu}\,{\nabla}_{\mu}\left\{ 
 \left(\rho\,h+{\|b\|}^2\right)\,U^{\mu}\,U^{\nu}+
 \left(P+{{\|b\|}^2 \over 2}\right){g}^{\mu\,\nu}-
 b^\mu\,b^\nu+\mathbf{R}^{\mu \nu}\right\} = 0 .
\end{equation}
By using  the law of baryon conservation
(\ref{barcons}), we recover after some algebra the local energy conservation law 
which now includes the projection of the divergence of the radiation tensor:
\begin{equation}\label{econs}
{\nabla}_{\mu} \left(\rho\,\epsilon\,U^{\mu}\right)+
 P\,{\nabla}_{\mu} U^{\mu}-U^\nu\,\nabla_\mu{\mathbf{R}^{\mu}}_\nu= 0.
\end{equation}
Applying the definition for the auxiliary energy function $E$, and also using 
\begin{equation}\label{tensoridentity}
{\Gamma_{\mu\,\nu}}^\gamma\,{\mathbf{\Pi}_\gamma}^\mu =
 {1 \over 2}\,\mathbf{\Pi}^{\gamma\,\mu}\,\partial_\nu\,g_{\mu\,\gamma} = 
-{1 \over 2}\,\mathbf{\Pi}_{\gamma\,\mu}\,\partial_\nu\,g^{\mu\,\gamma} 
\end{equation}
in expanding the radiation term,
the energy equation is rewritten as follows:
\begin{eqnarray} \label{enfinal}
 \partial_{t}\left(E\right)+{1 \over \sqrt{\gamma}}\,
\partial_{i}\left(\sqrt{\gamma}\,E\,V^i\right)
+ P\,\partial_{t}\left(W\right) + 
{P\over\sqrt{\gamma}}\,\partial_{i}\left(\sqrt{\gamma}\,W\,V^i\right) -
{W\over \alpha}\partial_t\,{\mathbf{R}^t}_t&-&\\\nonumber
{W\over \alpha}V^i\partial_t\,{\mathbf{R}^t}_i-
{W\over \alpha^2\sqrt{\gamma}}\partial_j\,(\alpha\sqrt{\gamma}{\mathbf{R}^j}_t)-
{W\over \alpha^2\sqrt{\gamma}}V^i\partial_j\,(\alpha\sqrt{\gamma}{\mathbf{R}^j}_i)-
{1\over 2}\mathbf{R}_{\mu \epsilon}{W\over \alpha}V^k\partial_k\,g^{\mu \epsilon}
&=& 0 ,
\end{eqnarray}
where the contributions from the various moments of the radiation field
significantly alter the structure of the equation, introducing additional transport
terms and, new to this portion of the code, metric derivatives. The  components of the
radiation tensor, ${\mathbf{R}^{\mu}}_\nu$ and $\mathbf{R}_{\mu \nu}$, have been left in their unsimplified form
 and are obtained from ${\mathbf{R}^{\mu \nu}}$ prescribed by the
diffusion approximation by lowering indices with the metric tensor.

The momentum conservation equations follow from applying the projection tensor
$h_{\mu \nu} = g_{\mu \nu}+U_\mu\,U_\nu$ to conservation law (\ref{tmncons}),
leading to $\nabla_\mu\,{\mathbf{T}^\mu}_\nu =0$. Expanding,
\begin{equation}\label{mom.1}
\nabla_\mu\,{\mathbf{T}^\mu}_\nu = {\nabla}_{\mu}\left\{ 
 \left(\rho\,h+{\|b\|}^2\right)\,U^{\mu}\,U_{\nu}+
 \left(P+{{\|b\|}^2 \over 2}\right){\delta^\mu}_\nu-
 b^\mu\,b_\nu+{\mathbf{R}^\mu}_\nu\right\} = 0 .
\end{equation}
Using the definition of momentum $S_\nu$ the first term in the
preceding expression can be rewritten as $S_\nu\,V^\mu/\alpha$ and  
simplified to ${S_\nu\,S^\mu / \alpha\,S^t}$, and applying  (\ref{divten})
and (\ref{tensoridentity}) 
to the radiation component, we rewrite the momentum
equation as
\begin{eqnarray}\label{mom.2}
{1 \over \alpha\,\sqrt{\gamma}}\,
\partial_\mu\,\sqrt{\gamma}\,S_\nu\,V^\mu +{1 \over 2\,\alpha}\,
{ S_\alpha\,S_\beta \over S^t }\,\partial_\nu\,g^{\alpha \beta}+
\partial_\nu\,\left(P+{{\|b\|}^2 \over 2}\right)-
{1 \over \alpha\,\sqrt{\gamma}}\,
\partial_\mu\,\alpha\,\sqrt{\gamma}\,b^\mu\,b_\nu&-&\\
\nonumber {1 \over 2}\,
 b_\alpha\,b_\beta\,\partial_\nu\,g^{\alpha \beta}+
\partial_t\,{\mathbf{R}^t}_\nu+
{1 \over \alpha \sqrt{\gamma}}\partial_i\,(\alpha \sqrt{\gamma}{\mathbf{R}^i}_\nu)+
{1 \over 2} {\mathbf{R}}^{\mu \epsilon} \partial_\nu\,(g_{\mu \epsilon})
&=& 0 .
\end{eqnarray}
To obtain the final form of the equations, multiply (\ref{mom.2})
by the lapse $\alpha$, split the $\mu$ index into
its space ($i$) and time ($t$) components, and restrict $\nu$ to the
spatial indices ($j$) only:
\begin{eqnarray}\label{mom.3}
\partial_t\left(S_j-\alpha\,b_j\,b^t+\alpha\,\partial_t\,{\mathbf{R}^t}_j\right)+
  {1 \over \sqrt{\gamma}}\,
  \partial_i\,\sqrt{\gamma}\,\left(S_j\,V^i-\alpha\,b_j\,b^i+\alpha\,\partial_t\,{\mathbf{R}^i}_j\right)&+&\\
 \nonumber  {1 \over 2}\,\left({S_\epsilon\,S_\mu \over S^t}-
  \alpha\,b_\mu\,b_\epsilon+\alpha\,{\mathbf{R}}_{\mu \epsilon}\right)\,
  \partial_j\,g^{\mu\,\epsilon}+
  \alpha\,\partial_j\left(P+{{\|b\|}^2 \over 2}\right) &=& 0 .
\end{eqnarray}

\newpage
\section{Constructing the Radiation Tensor\label{radtensor}}

The radiation tensor is most easily
computed in the reference frame that is locally comoving with the fluid (MM84).
Since the rGRMHD code uses Boyer-Linquist coordinates, it is necessary to introduce
transformations that translate between the two reference frames. This is done in two stages,
first by transforming the Boyer-Linquist coordinate basis to its corresponding orthonormal
basis (i.e. the ZAMO or LNRF frame; see Misner, Thorne \& Wheeler, 1973), and then applying a Lorentz boost from this frame to the comoving frame.

In the Boyer-Lindquist frame, denote the radiation tensor with plain indices, $\mathbf{R}^{\mu \nu}$;
in the orthonormal frame, denote it with carets, $\mathbf{R}^{\widehat{\mu} \widehat{\nu}}$; and
in the comoving frame, denote it with tildes, $\mathbf{R}^{\widetilde{\mu} \widetilde{\nu}}$.
The transformation from the Boyer-Lindquist frame to the orthonormal (ZAMO) frame is accomplished
by the Lorentz transformation ${\Lambda^{\widehat{\mu}}}_{\mu}$, and the inverse transformation
 ${\Lambda^{\mu}}_{\widehat{\mu}}$ transforms in the opposite direction. The
transformation
from the orthonormal basis to the comoving frame is accomplished by the boost 
${L^{\widetilde{\mu}}}_{\widehat{\mu}}$
and the boost from the comoving frame to the orthonormal frame,
${L^{\widehat{\mu}}}_{\widetilde{\mu}}$.
In each case, it is understood that the argument to the boost is  the 4-velocity in the orthonormal frame.

Once the radiation tensor has been computed in the comoving frame (see Appendix E), it
needs to be transformed to the Boyer-Lindquist frame by first boosting ``down'' 
to the orthonormal frame, 
\begin{equation}
\mathbf{R}^{\widehat{\mu} \widehat{\nu}} = {L^{\widehat{\mu}}}_{\widetilde{\mu}}\,\mathbf{R}^{\widetilde{\mu} \widetilde{\nu}}\,\left({L^{\widehat{\nu}}}_{\widetilde{\nu}}\right)^T
\end{equation}
followed by the coordinate transformation to Boyer-Lindquist,
\begin{equation}
\mathbf{R}^{\mu \nu} = {\Lambda^{\mu}}_{\widehat{\mu}}\,\mathbf{R}^{\widehat{\mu} \widehat{\nu}}\,
\left( {\Lambda^{\nu}}_{\widehat{\nu}}\right)^T
\end{equation}
It is also possible to combine these transformations,
\begin{equation}
\mathbf{R}^{\mu \nu} = \left( {\Lambda^{\mu}}_{\widehat{\mu}}\,{L^{\widehat{\mu}}}_{\widetilde{\mu}}\right)\,\mathbf{R}^{\widetilde{\mu} \widetilde{\nu}}\,\left({
 {\Lambda^{\nu}}_{\widehat{\nu}}\,L^{\widehat{\nu}}}_{\widetilde{\nu}}\right)^T.
\end{equation}
The expressions for the composite boost/basis transformations 
are given in T07.

In order to perform the boost, it is necessary to first form the 4-velocity in the Boyer-Lindquist
frame from the code variables for the transport velocity $V^i$ and Lorentz factor, $W$, and then
transform this 4-vector to the orthonormal frame
\begin{equation}
{U}^{\widehat{\mu}} =  {\Lambda^{\widehat{\mu}}}_{\mu}\,{U}^{\mu}.
\end{equation}
It is ${U}^{\widehat{\mu}}$ that sets the parameters of the boost transformation. 

\newpage
\section{The Diffusion Approximation\label{diffusion}}

At this intermediate stage of development of the rGRMHD code, it is possible to test the new
components without having need of the numerical implementation of the radiative transfer equation (RTE).
This is done by working in the first-order diffusion approximation (MM84) where the radiation tensor
can be constructed directly from the local temperature and density of the fluid. The 
comoving frame radiation tensor is given by
\begin{equation}
\mathbf{R}^{\widetilde{\mu} \widetilde{\nu}} =
\left[
\begin{array}{cc}
  a_{\rm R}\,T^4& -{K_{\rm R} \over c}\nabla T  \\
 & \\
  -{K_{\rm R} \over c}\nabla T& {1 \over 3} a_{\rm R}\,T^4\,{\bf I}   
\end{array}
\right]
\end{equation}
where $  a_{\rm R} = 7.57 \times 10^{-15}\,{\rm erg\,cm^{-3}\,K^{-4}}$ is the radiation constant
and $K_{\rm R}$ is the radiative conductivity, given by 
\begin{equation}
K_{\rm R} = {4 \over 3}\,c\,\lambda_{\rm R}\,a_{\rm R}\,T^3,
\end{equation}
where $\lambda_{\rm R} = 1/\chi_{\rm R}$ is the Rosseland mean free path, $\chi_{\rm R}$
the Rosseland mean opacity. For the range of densities and temperatures encountered in the simulated accreting fluid, the Rosseland mean opacity, $\chi_{\rm R}$,
can be approximated using the results of LCS91 who provide a set 
analytic expressions for the 
mean opacity of a zero-metallicity gas for a range of values that overlap the physical scenario 
considered here. (See Appendix \ref{units} for a more extensive discussion of the
scaling of code variables to astrophysical units.)

The temperature gradient, $\nabla T \equiv T_{,\,\widetilde{i}}$ , that appears in the radiative flux must be
computed in the comoving frame by transforming simulation data from the
Boyer-Lindquist frame. From the results of the previous appendix, the
gradient of any scalar quantity in the Boyer-Lindquist frame is evaluated in the comoving frame as follows:
\begin{equation}
\mathcal{X}_{,\,\widetilde{\mu}} = ({L^{\widehat{\mu}}}_{\widetilde{\mu}})^T\,
( {\Lambda^{\mu}}_{\widehat{\mu}})^T\,\mathcal{X}_{,\,\mu}
\end{equation}

\section{Extended Diagnostics\label{diagnostics}}

The set of evolution diagnostics described in DHK has been augmented
to monitor the radiative components.
The complete dumps of the code variables that are saved at
regular intervals now include all components of the radiation tensor.
The shell-averaged diagnostics, $\langle{\cal X}\rangle(r,t)$, are defined as
\begin{equation}\label{avgdef}
\langle{\cal X}\rangle(r,t) = {1 \over {{\cal A}}(r)} \int\int{ 
 {\cal X}\,\sqrt{-g}\, d \theta\,d \phi}
\end{equation}
where the area of a shell is ${\cal{A}}(r)$ and the bounds of integration 
cover the full range of the $\theta$ and $\phi$ grids.  
The rGRMHD code now includes shell-averaged values of radiative energy density,
$\langle\mathcal{E}\rangle$.
Fluxes through the shell are computed in a similar manner, but
not normalized with the area;  these diagnostics now include the radial components
$\langle \mathcal{F}^r\rangle$, $\langle \mathcal{P}^{rr}\rangle$, $\langle \mathcal{P}^{r \theta}\rangle$, 
and $\langle \mathcal{P}^{r \phi}\rangle$.

Volume-integrated quantities are computed using
\begin{equation}\label{3avgdef}
\left[{\cal Q}\right] = \int\int\int{ 
 {\cal Q}\,\sqrt{-g}\, dr\,d \theta\,d \phi}.
\end{equation}
The volume-integrated quantities computed and saved as a function of
time are the total rest mass, $\left[ \rho
U^t \right]$, angular momentum $\left[ {\mathbf{T}^t}_{\phi}\right]$, total
energy $\left[ {\mathbf{T}^t}_{t}\right]$ (exclusive of radiation), and  total
radiative energy
$\left[ {\mathbf{R}^t}_{t}\right]$. 

In addition, the history mechanism now optionally computes and dumps ray-traced images 
of simulation data by invoking a ray tracing component, as discussed in Appendix \ref{raytracing}.

\section{Ray Tracing\label{raytracing}}

The simulations that can be envisaged with both the GRMHD and rGRMHD codes
provide ever-increasing realism. In order to reconcile the results of such simulations
with the growing body of high-resolution observations  in various parts of the electromagnetic spectrum from earth- and space-based
observatories, the production of
simulated emission maps is a matter of increasing importance. To this end, a ray tracing
component has been added to the rGRMHD code's history mechanism that
is capable of producing ray traced images of simulated observables as would
be seen by a detector (pixel array) at a large distance from the central black hole. This is
accomplished by solving the geodesic equations in the Kerr metric for
each pixel in this array; the manner in which this is done largely follows Reynolds {\it et al.} (1999), but
the rGRMHD code uses analytic expressions for the turning points of the ODE describing the
polar coordinate of the polar angle $\Theta$. These solutions are solved once and coded as
Chebyshev polynomials, which allows for efficient archiving and fast computation (see Press {\it et al.}, 1992,  for the approach, though the rGRMHD code
implements the compression and retrieval routines differently).
Four Chebyshev polynomials are stored for each pixel, one for each Boyer-Linquist component of a
photon trajectory; i.e. $T(\lambda)$,  $R(\lambda)$,  $\Theta(\lambda)$,  and $\Phi(\lambda)$,
where $\lambda$ is an affine parameter common to all four polynomials for each geodesic. The
polynomials are interrogated through this common parameter as part of the history calculations to
map a particular emission event in the simulation data to a particular
pixel in the image plane. Photons emitted from different regions of
the simulation volume do not arrive at the pixel plane at the same time, so the code
maintains a stack of pixel planes, and maps arrival times to a particular pixel array in this
stack. When a
given pixel array in the stack slips out of causal contact with the simulation data
it is flushed to memory as a completed image. The array in question is cleared
and placed at the far end of the pixel stack to receive data from the portions of
the simulation volume just coming into causal contact with the pixel stack. This
ring buffer structure keeps memory usage to a minimum.
The pixel stacks produced by this tool allow the production
of static (time-averaged or instantaneous) and animated emission maps. 

Ray tracing stops at unit optical depth. Currently, no attempt is made
to account for refraction or reflection of photons on their flight to the
pixel plane. 

\begin{figure}[htbp]
\begin{center}
\includegraphics[width=3.5in]{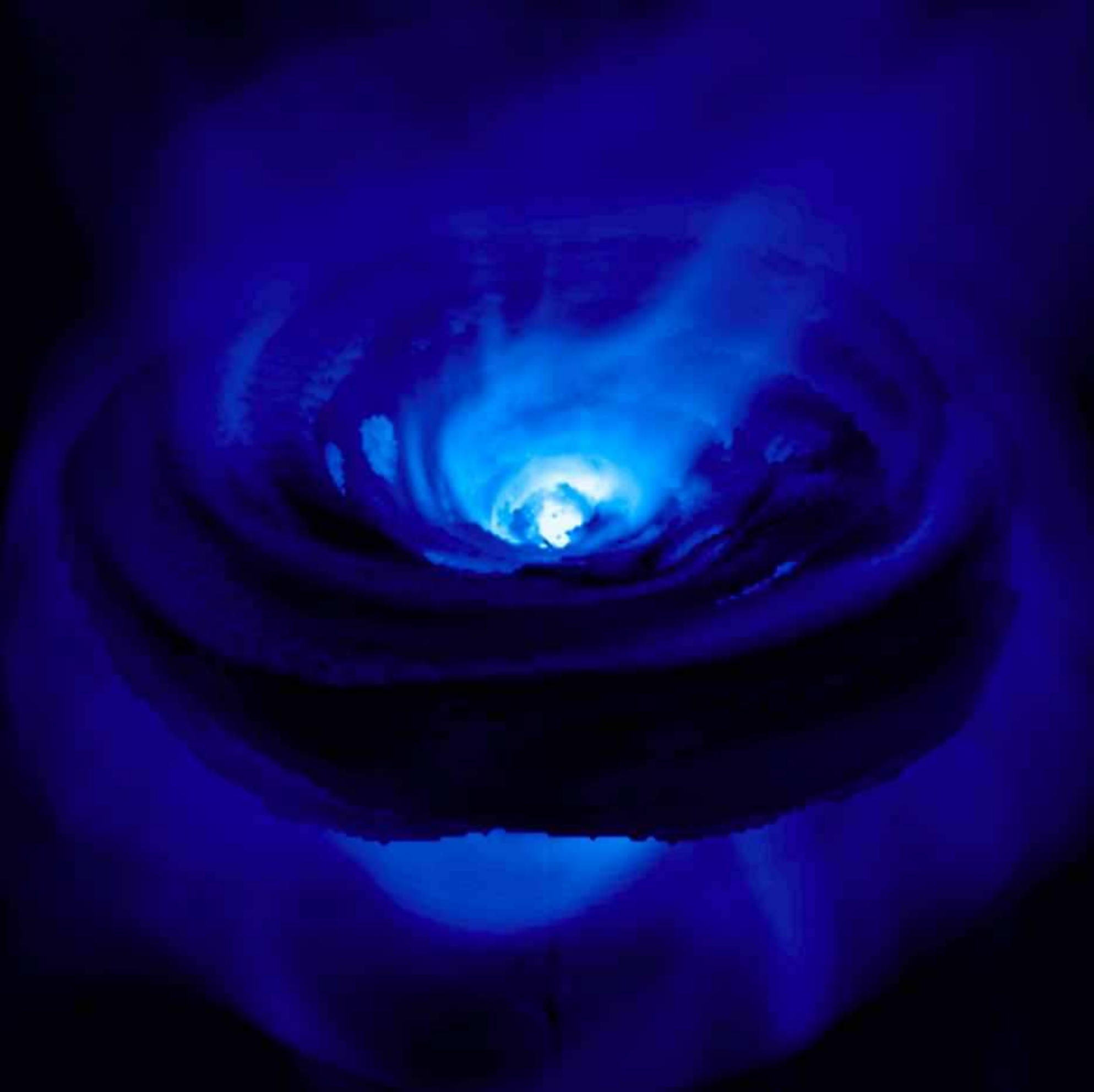}
\caption{Ray Traced Image of Poynting Flux in 3D GRMHD Simulation}
\label{Poynt3D}
\end{center}
\end{figure}

Several types of emission map can be constructed by the ray tracer, depending on the needs of a particular simulation. 
Since the full range of code variables can be accessed by the history mechanism,
the range of possible maps that can be produced is quite broad. For the simulations described in this paper, ray tracing was not used.
In part, this is due to the fact that axisymmetric simulations are not conducive to `realistic'
ray tracing, even though the ray tracer does render images by wrapping observables 
around the full azimuth. 
To better see the capabilities of the ray tracer, Figure \ref{Poynt3D} shows a test
image from an unpublished 3D simulation analogous to model KDP of DHK (main difference, $\Gamma = 4/3$
instead of $5/3$). The fully ray traced image is a projection of Poynting flux,  onto the
geodesic's tangent vector, ${{\mathbf{T}^t}_\mu}_{(EM)}\,\cal{X}^\mu$. Though various
emission maps have been tested, the Poyting flux maps remain the most appealing
for simple demonstrations in that they capture in a single image the cool disk, the hot plunging flow near the
black hole, and the wispy traces of the funnel outflow and jet. Ray traced images from
axisymmetric simulations are far less appealing due to pervasive artifacts due
to the axisymmetric grid, and are not reproduced here.

The rGRMHD code is an MPI-based parallel application that uses
a form of domain decomposition to achieve good performance on parallel
platforms. Introducing ray tracing in this context could adversely
affect performance because the process of tracing photons through the simulation volume
could introduce severe performance bottlenecks. To maintain performance, a client-server model was
adopted where many ray tracing functions are handled by a processor
dedicated to this task, freeing the remaining processors to
carry out the task of evolving the simulation. Some computational burden remains
with the simulation processors as part of the existing history mechanism, but this
burden is well balanced and does not degrade parallel performance (i.e. scalability), though it does add considerably to the turn-around time for a simulation so that
ray tracing remains a very expensive optional feature for simulations.

\newpage
\section{Calibration and Unit Conversion\label{units}}

The rGRMHD code uses relativistic units ($G=c=1$), with a central black hole taken to have unit 
mass ($M_{\rm bh}=1$). The opacity functions that are used during the construction of the radiation
tensor are expressed in cgs units, and are functions of the density and temperature of the gas, which are
evaluated in code units; to express the radiation tensor in code units, it is therefore
necessary to
scale the code units to physical units (cgs), and vice-versa. Furthermore,
the test fluid approximation which is used in the GRMHD numerical scheme effectively
decouples the mass of the accreting fluid from that of the central black hole, so that the initial
state can be assigned an arbitrary density, along with an internal energy that is scaled in relation
to this density to distinguish between ``hot'' and ``cold'' initial tori. However, when a radiative component
is introduced, this arbitrary scaling is no longer appropriate since radiative effects must
be tied to the mass/energy and time scales of the central black hole. Therefore, both the initial state and
the evolving fluid variables must be calibrated to an astrophysical scenario to correctly
capture radiative effects.

The initial state consists of a torus with an inner edge
at $r_{in}=6.8\,M_{bh}$, an outer edge at $r_{out}\approx 30\,M_{bh}$ and a scale height of $h \approx 6.1\,M_{bh}$, measured on the $0.01\,\epsilon_{max}$ contour at the
pressure maximum. In order to match
this initial state to an astrophysical scenario, assume that the torus contains 100 solar masses of gas at a temperature
of $\sim 100$ K, and that it
orbits a supermassive black hole $10^8$ solar masses.
 
For this choice of parameters, we derive a mean physical density $\overline{\rho}_{\rm (cgs)}$ which can be compared during initialization to the grid-averaged density in code
units. The density scaling parameter is set accordingly as 
\begin{equation}
\eta={\overline{\rho}_{\rm (cgs)} \over \overline{\rho}_{\rm (code)}}
\end{equation}
and is subsequently used to convert the density variable during simulations.

The temperature scaling parameter 
\begin{equation}
\xi={{T}_{\rm (cgs)} \over {T}_{\rm (code)}}
\end{equation}
converts between the two sets of units during the simulation, where ${T}_{\rm code} \equiv (\Gamma -1)\,\epsilon=  (\Gamma -1)\,E/D$. This parameter is also set from the initial conditions by
computing the ratio of the mean physical temperature (a model-dependent parameter) and the grid-averaged temperature in code units of the initial torus.

The scaling parameters $\eta$ and $\xi$ are used to convert density and temperature 
during the computation of the Rosseland mean opacity, $\chi_{\rm R\,(cgs)}$, using the 
approximation supplied by LCS91. Once $\chi_{\rm R\,(cgs)}$ has been calculated, 
it is converted to code units and used to construct the radiation tensor in the comoving frame. 

To construct $\mathbf{R}^{\mu \nu}$, we need to express the radiation constant, $a_{\rm R}$, and
the radiative conductivity, $K_{\rm R}$, in code units. The radiation constant can be set during 
initialization by relating initial values of the dimensionless ratio of the time-time components
of the fluid energy momentum tensor and the radiation tensor,
\begin{equation}
{\mathbf{T}^{tt}_{\rm (fluid)} \over \mathbf{R}^{tt}}
\end{equation}
which must remain unchanged in converting from cgs to code units. 
In general,
\begin{equation}
\mathbf{T}^{tt}_{\rm (fluid)} = \rho\,h\,(U^t)^2 +g^{tt}\,P=-{g^{tt} \over W}\left[{(D + \Gamma\,E)\,W^2 - (\Gamma-1)\,E}\right]
\end{equation}
This expression can be simplified considerably, however, since $\mathbf{T}^{tt}_{\rm (fluid)}$
is to be evaluated for the initial torus, for which $W\approx 1$, and $g^{tt} \approx -1$ in
the vicinity of the disk pressure maximum. To a very good approximation, $\mathbf{T}^{tt}_{\rm (fluid)}$
can be approximated by the rest-energy of the initial torus, so
$\mathbf{T}^{tt}_{\rm (fluid)\,(code) } \approx \rho_{\rm (code)}$ and $
\mathbf{T}^{tt}_{\rm (fluid)\,(cgs) } \approx \rho_{\rm (cgs)}\,c^2$.
With the rest energy scaling parameter
\begin{equation}
\zeta = {\rho_{\rm (cgs)}\,c^2 \over \rho_{\rm (code)} }= \eta\, c^2
\end{equation}
it follows that the dimensionless ratio can be approximated as follows:
\begin{equation}
{\mathbf{T}^{tt}_{\rm (fluid)} \over \mathbf{R}^{tt}} \rightarrow 
{\rho_{\rm (code)} \over a_{\rm R\,(code)}\,T_{\rm (code)}^4} = 
{\rho_{\rm (cgs)} \,c^2 \over a_{\rm R\,(cgs)}\,T_{\rm (cgs)}^4} ,
\end{equation}
and the radiation constant $a_{\rm R\,(code)}$ can be obtained
\begin{equation}
a_{\rm R\,(code)}= a_{\rm R\,(cgs)}\,{\xi^4 \over \zeta}
\end{equation}

The radiative conductivity is given by (MM84)
\begin{equation}
K_{\rm R} = {4 \over 3}\,c\,\lambda_{\rm R}\,a_{\rm R}\,T^3,
\end{equation}
where $\lambda_{\rm R} = 1/\chi_{\rm R}$ is the Rosseland mean free path.
The Rosseland mean free path (units $cm^{-1}$) is converted to code units by using the
length-scale conversion
\begin{equation}
r_{\rm g (code)} = 1 \rightarrow r_{\rm g (cgs)}={{G}_{\rm (cgs)} \,M_{\rm bh\,(cgs)}\over {c}_{\rm (cgs)}^2}
\end{equation}

\newpage
\section{Initial State - Torus with Thermal Radiation \label{InitState}}

This Appendix reviews the initialization used in the earlier DHK03 simulations and discusses how the
initial state needs to be modified to study radiative effects.

\subsection{Initial Torus - Fluid Variables}

The initial state consists of  a
thick torus with a nearly-Keplerian distribution of angular
momentum, with specific angular momentum given by
$l=-U_\phi/U_t$ and angular velocity $\Omega = U^\phi/U^t$.
DHK03 provides a detailed derivation of the following analytic expression for
the
internal energy of the disk,
\begin{equation}\label{kd.11}
  \epsilon(r,\theta) = {1 \over \Gamma} \left(
  { U_{in} f(l_{in}) \over U_{t}(r,\theta) f(l(r,\theta))}-1\right) .
\end{equation}
where $U_{t}$ represents the binding energy, $U_{in}$ the surface 
binding energy, and $f(l) = {\|1 - k\,l^{\alpha+1}\|}^{1/(\alpha+1)}$. The
parameter $\alpha$ is related to the disk structural parameter $q$ by
 $\alpha = q/(q-2)$.
Also, $k=\eta^{-2/(q-2)}$ and $\eta$ is a constant used in constructing the
power-law rotation for the initial torus, 
\begin{equation}\label{kd.1} 
\Omega = \eta\,\lambda^{-q} 
\end{equation} 
where  $\lambda$ is given by
\begin{equation}\label{kd.7}
\lambda^2 = {l \over \Omega} 
 = l {\left(g^{t\,t}-l\,g^{t\,\phi} \right) \over 
      \left(g^{t\,\phi}-l\,g^{\phi\,\phi} \right)} .
\end{equation}

For a constant entropy adiabatic gas the pressure is given by $P =
\rho\,\epsilon\,(\Gamma - 1) = K\,\rho^\Gamma$, and density is given by
$\rho={\left[{\epsilon\,(\Gamma - 1) / K}\right]}^{1/(\Gamma - 1)}$.

A particular disk solution is specified by choosing the parameter $q$,
the entropy parameter $K$, and the angular momentum $l_{in}$ at
$r_{in}$, the inner edge of the disk.  For all simulations, the
location of the inner edge, $r_{in}=6.80\,M_{\rm bh}$, as well as parameters 
$K=0.01$, $q=1.66$, and $l_{in}=3.34$ are kept fixed.

\subsection{Initial Torus - Magnetic Field}

The initial magnetic field is obtained from the definition of $F_{\mu
\nu}$ in terms of the $4$-vector potential, $A_\mu$,
$F_{\mu \nu} = \partial_\mu A_{\nu} - \partial_\nu A_{\mu}$ .
Our initial field consists of axisymmetric poloidal field loops, laid
down along isodensity surfaces within the torus by defining
$A_{\mu} = (A_t,0,0,A_\phi)$, where
\begin{equation}\label{vecpot}
A_\phi = 
\cases{
k (\rho-\rho_{cut}) & for $\rho \ge \rho_{cut}$ \cr
0 & for $\rho < \rho_{cut}$},
\end{equation}
where $\rho_{cut}$ is a cutoff density corresponding to a particular
isodensity surface within the torus.  Using the above definition, it
follows that ${\cal{B}}^r = -\partial_\theta A_{\phi}$ and
${\cal{B}}^\theta = \partial_r A_{\phi}$.  The constant $k$ is set by
the input parameter $\beta$, the ratio of the gas pressure to the
magnetic pressure, using the volume-integrated gas pressure divided by
the volume-integrated magnetic energy density in the initial torus.  We
use $\beta=100$ in all runs.
The constant $\rho_{cut}$ is chosen to keep the initial magnetic field
away from the outer edge of the disk.  Here we use $\rho_{cut} = 0.5
\rho_{max}$, where $\rho_{max}$ is the maximum density at the center of
the torus, to ensure that the initial field loops are confined well
inside the torus.

\subsection{Initial Torus - External Medium}

The region outside the torus is initialized to a numerical vacuum that
consists of a cold, tenuous, non-rotating, unmagnetized gas. The
auxiliary density variable, $D$, in the vacuum is set to seven orders
of magnitude below the maximum value of $D$ in the  initial torus.
Similarly, the auxiliary energy variable $E$ is set ten orders of
magnitude below the initial maximum of $E$. These values define the
numerical floor of the code, below which $D$ and $E$ are not allowed to
drop. In practice, the numerical floor is rarely asserted during a
simulation, since outflow from the evolving torus quickly populates
the grid with a gas that, though of low density, lies above the
numerical floor. Further details on the numerical floor are given in
DH03.

\subsection{Initial Torus - Radiation Tensor}

Once the hydrodynamic variables have
been initialized, the radiation tensor in the diffusion approximation
is set using the temperature and density distributions. The radiation
tensor is computed from the initial state using the same routines that
compute its values during the simulation. The following procedure is
carried out in each computational zone:
\begin{enumerate}
\item density and temperature converted to cgs units (Appendix \ref{units})
\item Rosseland mean opacity computed from these values (Appendix \ref{diffusion})
\item Rosseland mean converted to code units (Appendix \ref{units})
\item comoving-frame radiation tensor computed from temperature and
opacity (Appendix \ref{diffusion}), and also from temperature gradient in
comoving frame  (Appendix \ref{radtensor})
\item comoving-frame radiation tensor converted to Boyer-Lindquist frame
 (Appendix \ref{radtensor})
\end{enumerate}


\begin{thebibliography}{99}
\bibitem{bh98} Balbus, S.~A. \& Hawley, J.~F.
 1998, Rev. Mod. Phys., 70, 1
\bibitem{DH} De Villiers, J.~P. \& 
 Hawley, J.~F. 2003, ApJ, 589, 458 (DH03)
\bibitem{DHK} De Villiers, J.~P., 
 Hawley, J.~F., \& Krolik, J.~H. 2003, ApJ, 599, 1238 (DHK03)
\bibitem{D:05} De Villiers, 
J.~P., Hawley, J.~F., Krolik, J.~H., \& Hirose, S. 2005, ApJ, 620, 878
\bibitem{D06} De Villiers, J.~P. 2006, astro-ph/0605744
\bibitem{FKR}
Frank, J., King, A., \& Raine, D. 2002,
Accretion Power in Astrophysics, 3rd ed. (Cambridge)
\bibitem{HSW1:84} 
 Hawley, J.~F., Smarr, L.~L., \& Wilson, J.~R., 1984, ApJ, 277, 296 (HSW)
\bibitem{PaperII} 
 Hirose, S., Krolik, J.~H., De Villiers, J.~P., \& Hawley, J.~F. 2004, ApJ, 
 606, 1083
\bibitem{HKS} 
 Hirose, S., Krolik, J.~H., \& Stone, J.~M. 2006, ApJ, 
 640, 901
\bibitem{KHH} 
 Krolik, J.~H., Hirose, S.,  \& Hawley, J.~F. 2005, ApJ, 
 620, 878
\bibitem{LCS} 
 Lenzuni, P., Chernoff, D.~F.,  \& Salpeter, E.~E. 1991, ApJS, 
 76, 758 (LCS91)
\bibitem{L66} 
 Lindquist, R.~W. 1966, Annals of Physics, 37, 487
\bibitem{mg04} Gammie, C.~F., McKinney, J.~C., \& Toth, G. 
 2003, ApJ, 589, 444
\bibitem{MM} Mihalas, D., \& Weibel-Mihalas, B. 1984. Foundations of Radiation Hydrodynamics (Mineola:Dover).
\bibitem{MTW} Misner, C.~W., Thorne,
 K.~S., \& Wheeler, J.~ A, 1973, Gravitation (San Francisco: W.H. Freeman)
\bibitem{M06} Mizuno, Y., Nishikawa, K.-I., Koide, S., Hardee, P., \& Fishman, G. J., 2006, http://arxiv.org/abs/astro-ph/0609004v1
\bibitem{NR}
Press, W.~H., Flannery, B.~P., Teukolsky, S.~A., \& Vetterling, W.~T. 1992,
Numerical Recipes in FORTRAN 77: The Art of Scientific Computing 
(Cambridge)
\bibitem{R98}
Reynolds, C.~S., Young, A.~I., Begelman, M.~C., \& Fabian, A.~C., 1998
ApJ, 514, 164
\bibitem{xx} 
 Schnittman, J.~D., Krolik, J.~H., \& Hawley, J.~F. 2006, ApJ, 
 651, 1031
\bibitem{Tak} 
 Takahashi, R., 2007, MNRAS, 382, 1041 (T07)
\end{thebibliography}
 \end{document}